# THE QUANTUM DYNAMICS OF THE COMPACTIFIED TRIGONOMETRIC RUIJSENAARS-SCHNEIDER MODEL

J. F. VAN DIEJEN AND L. VINET

ABSTRACT. We quantize a compactified version of the trigonometric Ruijsenaars-Schneider particle model with a phase space that is symplectomorphic to the complex projective space $\mathbb{CP}^N$. The quantum Hamiltonian is realized as a discrete difference operator acting in a finite-dimensional Hilbert space of complex functions with support in a finite uniform lattice over a convex polytope (viz., a restricted Weyl alcove with walls having a thickness proportional to the coupling parameter). We solve the corresponding finite-dimensional (bispectral) eigenvalue problem in terms of discretized Macdonald polynomials with $q$ (and $t$) on the unit circle. The normalization of the wave functions is determined using a terminating version of a recent summation formula due to Aomoto, Ito and Macdonald. The resulting eigenfunction transform determines a discrete Fourier-type involution in the Hilbert space of lattice functions. This is in correspondence with Ruijsenaars' observation that—at the classical level—the action-angle transformation defines an (anti)symplectic involution of $\mathbb{CP}^N$. From the perspective of algebraic combinatorics, our results give rise to a novel system of bilinear summation identities for the Macdonald symmetric functions.

## Contents

*Date*: September 1997.

Work supported in part by the Natural Sciences and Engineering Research Council (NSERC) of Canada, by le Fonds pour la Formation de Chercheurs et l'Aide à la Recherche (FCAR) du Québec, and at MSRI by NSF grant #DMS 9022140.

1. INTRODUCTION

In a recently published work [R3], Ruijsenaars presented a detailed study of the dynamics of the classical Sutherland-Moser particle model [Mo] and its "relativistic" deformation the trigonometric Ruijsenaars-Schneider model [RS, R1]. In addition, he also considered a closely related integrable system characterized by an $(N+1)$-particle Hamiltonian of the form

$$(1.1) \qquad H = \sum_{1\leq j\leq N+1} \cos(\beta p_j) \prod_{1\leq k\leq N+1,\, k\neq j} \left(1 - \frac{\sin^2(\frac{\alpha\beta g}{2})}{\sin^2 \frac{\alpha}{2}(x_j - x_k)}\right)^{1/2}.$$

The Hamiltonian in (1.1) differs from the standard trigonometric Ruijsenaars-Schneider Hamiltonian by the substitution $\beta \to i\beta$ (where $i = \sqrt{-1}$). Even though over the complex field both systems are equivalent, it turns out that their real (i.e. physical) dynamics are quite distinct. (Throughout we are assuming that our variables $x_j$, $p_j$ as well as the scale factors $\alpha, \beta$ and the coupling parameter $g$ are real-valued.) The main point is that $H$ (1.1) is periodic not only in the **x** but also in the **p** variables. This periodicity naturally prompts one to employ a phase space which—upon restricting attention to the relative motion in the center-of-mass frame—is bounded (in fact compact after a suitable completion). This is in contrast to the situation for the standard trigonometric Ruijsenaars-Schneider model (with $\cos(\beta p_j) \to \cosh(\beta p_j)$ and $-\sin^2(\alpha\beta g/2) \to +\sinh^2(\alpha\beta g/2)$ substituted in (1.1)), where the phase space is given by the (manifestly noncompact) cotangent bundle over the configuration space.

From now on the system determined by the Hamiltonian $H$ (1.1) will be referred to as the *compactified* trigonometric Ruijsenaars-Schneider model. It is the purpose of the present paper to investigate the corresponding quantum system. We will see that, in accordance with physical intuition, the Hilbert space for the quantum model becomes finite-dimensional. In essence, the Hilbert space in question consists of the space of all complex functions with support in a finite uniform lattice (grid) over classical configuration space. This configuration space has the geometry of a convex polytope consisting of a restricted Weyl alcove with walls that have a thickness determined by the value of the coupling parameter $g$. Matching the lattice so as to let it fit precisely over the configuration space, including the vertices (corner points) of the polytope, produces a quantization condition on $g$ that relates the coupling parameter to the size of the lattice. The quantum Hamiltonian is in turn given by a discrete difference operator with a step size that is equal to the distance between neighboring lattice points. Mathematically, our quantization condition on $g$ translates in vanishing conditions for the coefficients at the boundary lattice points, therewith guaranteeing that the discrete difference operator Hamiltonian is well-defined and self-adjoint as an operator in the Hilbert space of complex functions over the finite lattice.

For the quantum version of the standard trigonometric Ruijsenaars-Schneider model it is well-known (see e.g. the introduction of [D2] or Sect. 7.6.2 of [R4]) that the eigenfunctions may be expressed as a product of a factorized (ground-state) wave function and Macdonald polynomials (with $0 < q < 1$) [M2, M3, M4]. Here also, in the case of the compactified Ruijsenaars-Schneider model, the eigenfunctions turn out to be similarly expressible in terms of Macdonald polynomials. In contrast to the standard situation, however, in the compactified/discrete context of the present paper the parameters $q$ and $t$ lie on the unit circle and the diagonalization of the model involves only a finite number of Macdonald polynomials (viz., precisely as many as the number of lattice points = the dimension of the Hilbert space). The symmetry relations for the Macdonald polynomials [Ko1, M4, EK] have as consequence that the discrete kernel for the finite-dimensional eigenfunction transform is symmetric. This reflects the fact that we are actually dealing with a multivariate finite-dimensional doubly discrete *bispectral problem* in the sense of Duistermaat and Grünbaum [DG, W, G]. More concretely, the discrete eigenfunction kernel satisfies the same discrete difference equations in the "spectral" variables as it does in the "spatial" variables. Combined with the unitarity, the symmetry of the kernel furthermore implies that the eigenfunction transform determines a discrete Fourier-type involution in the Hilbert space of lattice functions. This is the quantum counterpart of the corresponding property of (the closure of) the action-angle transformation for the classical compactified Ruijsenaars-Schneider model, which turns out to define an *involutive* (anti)symplectomorphism of the classical phase space ($\cong \mathbb{CP}^N$) [R3].

The paper is organized as follows. We first recall in Sect. 2 some properties of the classical compactified Ruijsenaars-Schneider system taken from [R3]. Specifically, we discuss the commuting integrals, the configuration space (viz. the restricted Weyl alcove with walls of thickness proportional to $g$) and also the phase space of the model. A rather remarkable property of the dynamical system under consideration is that the phase space for the relative particle motion in the center-of-mass frame becomes, after a suitable compactification, isomorphic to the complex projective space $\mathbb{CP}^N$. In particular, globally the compactified phase space *does not* have a topology of product form (it is *not* topologically equivalent to the direct product of the configuration space times a (real) $N$-dimensional torus). Sect. 3 goes on to demonstrate how canonical quantization ($p_j \to \partial/i\partial x_j$) gives rise to discrete difference operator Hamiltonians acting in a finite-dimensional Hilbert space of lattice functions over the classical configuration space. In Sect. 4, the spectrum of the quantum model is determined in explicit form and the corresponding wave functions are expressed in terms of Macdonald polynomials with $q, t$ on the unit circle. In order to normalize our wave functions such that their $L^2$-norms are equal to one, it is necessary to evaluate a terminating version of a recently found summation formula due to Aomoto, Ito and Macdonald [Ao, I, M5]. The details explaining how to truncate the Aomoto-Ito-Macdonald sum so as to arrive at its terminating version are relegated to the first of two appendices at the end of the paper (Appendix A). In a second appendix (viz. Appendix B), some useful properties of the Macdonald symmetric functions taken from [M2, M4] have been collected. These properties were needed in Sect. 4 for the diagonalization of the quantum model. We have also taken the opportunity to reformulate here some of our results from the viewpoint of algebraic combinatorics. This leads us, in particular, to a new system of bilinear summation identities for the Macdonald symmetric functions (cf. Proposition B.2). The paper closes in Sect. 5 with some miscellaneous results and remarks. Among other things, it is pointed out that: (i) the results on the eigenfunctions give rise to a Discrete Fourier-type Transform for lattice functions over the restricted Weyl alcove, (ii) both the maximal and the minimal energy of the compactified Ruijsenaars-Schneider model are at the quantum level the same as at the classical level (the quantization discretizes the energy levels but does not shift the spectrum) and (iii) the dimension of our Hilbert space is in agreement with the dimension predicted by the Riemann-Roch-Hirzebruch formula for $\mathbb{CP}^N$, in the framework of geometric quantization [HK, Si, Hu].

*Note:* By means of a symplectic (with respect to the standard symplectic form $\sum_j dx_j \wedge dp_j$) rescaling $(\mathbf{x}, \mathbf{p}) \to (\beta \mathbf{x}, \beta^{-1} \mathbf{p})$ one absorbs the scale parameter $\beta$ in $\alpha$ (cf. (1.1)). In this paper we will from now on pick $\beta = 1$ without loss of generality. At the quantum level this means that we have scaled our variables such that the step size of the discrete difference operator Hamiltonians becomes equal to one (or to $\sqrt{\frac{N}{N+1}}$ after projection onto the center-of-mass hyperplane), cf. Sect. 3.1 and the remark at the end of Sect. 3.

## 2. The classical system

This section serves to summarize some of the basic properties of the classical compactified Ruijsenaars-Schneider model that are discussed in more detail in [R3]. Since we are primarily interested in the relative particle motion in the center-of-mass frame, it will be convenient to employ root system notation. For our purposes it suffices to restrict attention to the root system of type $A_N$. Some relevant preliminaries have been collected in the first subsection. For further information regarding root systems the reader is referred e.g. to (the Planches in) Bourbaki [B].

### 2.1. Some notational preliminaries.
Below the vectors $\mathbf{e}_1, \ldots, \mathbf{e}_{N+1}$ always represent the unit vectors constituting the standard basis of $\mathbb{R}^{N+1}$ and $\langle \cdot, \cdot \rangle$ denotes the (usual) inner product with respect to which this standard basis becomes an orthonormal basis (i.e. $\langle \mathbf{e}_j, \mathbf{e}_k \rangle = \delta_{j,k}$).

Let $E$ be the center-of-mass hyperplane

$$(2.1) \quad E = \{\mathbf{x} \in \mathbb{R}^{N+1} \mid x_1 + \cdots + x_{N+1} = 0\}.$$

A natural basis $\{\mathbf{a}_1, \ldots, \mathbf{a}_N\}$ for $E$ is given by the *simple roots*

$$(2.2) \quad \mathbf{a}_j = \mathbf{e}_j - \mathbf{e}_{j+1}, \quad j = 1, \ldots, N.$$

The associated dual basis $\{\omega_1, \ldots, \omega_N\}$—determined uniquely by the property that $\langle \omega_j, \mathbf{a}_k \rangle = \delta_{j,k}$—is realized explicitly by the *fundamental weights*

$$\omega_j = (\mathbf{e}_1 + \cdots + \mathbf{e}_j) - \frac{j}{N+1}(\mathbf{e}_1 + \cdots + \mathbf{e}_{N+1}), \quad j = 1, \ldots, N. \tag{2.3}$$

To these two bases of $E$ (2.1) one can associate the *root lattice*

$$Q = \mathbb{Z} - \text{Span}\{\mathbf{a}_1, \ldots, \mathbf{a}_N\} \tag{2.4}$$

and the *weight lattice*

$$\Lambda = \mathbb{Z} - \text{Span}\{\omega_1, \ldots, \omega_N\} \tag{2.5}$$

as well as the corresponding positive semi-lattices (or integral cones)

$$Q^+ = \mathbb{N} - \text{Span}\{\mathbf{a}_1, \ldots, \mathbf{a}_N\} \tag{2.6}$$

and

$$\Lambda^+ = \mathbb{N} - \text{Span}\{\omega_1, \ldots, \omega_N\}, \tag{2.7}$$

respectively (where in our conventions the set of natural numbers $\mathbb{N}$ does include the number zero). The semi-lattice $\Lambda^+$ (2.7) is usually referred to as the cone of *dominant weights*. This cone is partially ordered by the *dominance order*, which is defined for $\lambda, \mu \in \Lambda^+$ by

$$\mu \preceq \lambda \quad \text{iff} \quad \lambda - \mu \in Q^+ \tag{2.8}$$

(and $\mu \prec \lambda$ iff $\mu \preceq \lambda$ and $\mu \neq \lambda$).

The *Weyl group* generated by the reflections in planes orthogonal to the simple roots $\mathbf{a}_1, \ldots, \mathbf{a}_N$ (2.2) is realized explicitly as the group of permutations $\sigma \in S_{N+1}$ acting on the vectors $\mathbf{e}_1, \ldots, \mathbf{e}_{N+1}$ by

$$\sigma(\mathbf{e}_j) := \mathbf{e}_{\sigma(j)}. \tag{2.9}$$

The (unique) orbit of the basis vectors $\mathbf{a}_j$ with respect to the $S_{N+1}$-action consists of the *roots*

$$A_N = \{\mathbf{e}_j - \mathbf{e}_k \mid 1 \leq j \neq k \leq N+1\}. \tag{2.10}$$

For future reference we also need to identify the *positive roots*

$$A_N^+ = A_N \cap Q^+ = \{\mathbf{e}_j - \mathbf{e}_k \mid 1 \leq j < k \leq N+1\}, \tag{2.11}$$

the *maximal root*

$$\mathbf{a}_{max} = \sum_{1 \leq j \leq N} \mathbf{a}_j = \mathbf{e}_1 - \mathbf{e}_{N+1} = \omega_1 + \omega_N \tag{2.12}$$

(this root is maximal in $A_N$ (2.10) with respect to the partial dominance order in (2.8)), and the *weighted half sum over the positive roots*

$$
\begin{aligned}
(2.13) \quad \rho &= \frac{g}{2} \sum_{\mathbf{a} \in A_N^+} \mathbf{a} = \frac{g}{2} \sum_{1 \leq j < k \leq N+1} (\mathbf{e}_j - \mathbf{e}_k) \\
&= \frac{g}{2} \sum_{1 \leq j \leq N+1} (N - 2(j-1)) \mathbf{e}_j \\
&= g(\omega_1 + \cdots + \omega_N).
\end{aligned}
$$

2.2. **Integrability.** The Hamiltonian $H$ (1.1) is known to be integrable: a complete set of integrals in involution is given explicitly by [RS, R1]

$$
(2.14) \quad H_r = \sum_{\substack{J \subset \{1,\ldots,N+1\} \\ |J|=r}} \cos(\textstyle\sum_{j \in J} p_j) \prod_{\substack{j \in J \\ k \notin J}} \left(1 - \frac{\sin^2(\frac{\alpha g}{2})}{\sin^2 \frac{\alpha}{2}(x_j - x_k)}\right)^{1/2}
$$

$r = 1, \ldots, N+1$. (Recall that we have rescaled the variables such that $\beta = 1$, cf. the note at the end of the introduction.) Observe that $H_r$ (2.14) specializes for $r = 1$ to the Hamiltonian $H$ (1.1) and that $H_{N+1} = \cos(p_1 + \cdots + p_{N+1})$, reflecting the translational invariance of the model. The projection of the $H_r$-flow onto the center-of-mass hyperplane $x_1 + \cdots + x_{N+1} = 0$ is governed by the reduced Hamiltonian $\mathcal{H}_r$, written conveniently in root system notation as

$$
(2.15) \quad \mathcal{H}_r = \sum_{\nu \in S_{N+1}(\omega_r)} \cos(\langle \nu, \mathbf{p} \rangle) \prod_{\substack{\mathbf{a} \in A_N \\ \langle \mathbf{a}, \nu \rangle = 1}} \left(1 - \frac{\sin^2(\frac{\alpha g}{2})}{\sin^2 \frac{\alpha}{2} \langle \mathbf{a}, \mathbf{x} \rangle}\right)^{1/2}
$$

$r = 1, \ldots, N$ (for $H_{N+1}$ the reduced flow in center-of-mass plane is of course trivially stationary). Here $\mathbf{x} := (x_1, \ldots, x_{N+1})$, $\mathbf{p} := (p_1, \ldots, p_{N+1})$ and the sum in (2.15) is over all weights $\nu \in \Lambda$ (2.5) that lie in the $S_{N+1}$-orbit (recall the action (2.9)) of the $r$-th fundamental weight vector $\omega_r$ (2.3).

2.3. **The reduced phase space for the relative particle motion: $\mathbb{CP}^N$.** Let us from now on assume that the scale factor $\alpha$ is positive and that the parameter $g$ lies in the interval

$$
(2.16) \quad 0 < g < \frac{2\pi}{(N+1)\alpha}.
$$

In order to arrive at real-valued Hamiltonians $\mathcal{H}_r$ (2.15), one is led to employ a configuration space in which the particle distances $|x_j - x_k|$ are bounded from below by $g$ ($> 0$) and from above by $2\pi/\alpha - g$ ($> 0$). This is realized by picking as configuration space the submanifold $\Sigma_g$ of the center-of-mass plane consisting of the points $\mathbf{x} \in E$ (2.1) satisfying the conditions

 (i) $\langle \mathbf{a}_j, \mathbf{x} \rangle > g$ for $j = 1, \ldots, N$;
 (ii) $\langle \mathbf{a}_{max}, \mathbf{x} \rangle < 2\pi/\alpha - g$

(where the vectors $\mathbf{a}_1, \ldots, \mathbf{a}_N$ denote the simple roots (2.2) and $\mathbf{a}_{max}$ is the maximal root (2.12)). The parameter restriction (2.16) ensures that the submanifold $\Sigma_g \subset E$ determined by (i), (ii) is nonempty (add the $N$ inequalities from (i) and use (2.12) to compare with (ii)). Furthermore, $\Sigma_g$ has the geometry of an open convex polytope consisting of an alcove with walls of thickness $g/\sqrt{2}$ inside the Weyl alcove $\Sigma_0$ (which corresponds to the limit $g \downarrow 0$). The open convex polytope (or open simplex) $\Sigma_g$ is completely determined by the $N+1$ vertices (corner points) $\rho$, $\rho+M\omega_r$ ($r = 1, \ldots, N$) with $M = \frac{2\pi}{\alpha} - (N+1)g > 0$. See Figure 1.

FIGURE 1. The restricted Weyl alcove with walls of thickness $g/\sqrt{2}$ for $N = 2$. The region of the inner alcove corresponds to the configurations spaces $\Sigma_g$ (without boundary) and $\overline{\Sigma}_g$ (with boundary) of the three-particle system in the center-of-mass hyperplane $x_1 + x_2 + x_3 = 0$. The vertices and boundary segments with/without the shifts between square brackets refer to the inner/outer triangle, respectively.

An obvious candidate for the phase space would now of course be the cotangent bundle over the configuration space: $T^*(\Sigma_g) \cong \Sigma_g \times E$. However, in view of the periodicity of the the Hamiltonians $\mathcal{H}_r$ (2.15) with respect to translations in $\mathbf{p}$ over vectors in the dilated root lattice $2\pi Q$ (cf. (2.4)), it is natural to restrict to a smaller phase space of the form $\Sigma_g \times T$, where $T$ is the $N$-dimensional torus $E/(2\pi Q)$. This torus can be coordinatized explicitly as

(2.17) $$T = \{\mathbf{p} \in E \mid -\pi < \langle \omega_r, \mathbf{p} \rangle \leq \pi, \ r = 1, \ldots, N\},$$

where the components $\langle \omega_r, \mathbf{p} \rangle$ of the vector $\mathbf{p}$ with respect to the basis of fundamental weights $\{\omega_1, \ldots, \omega_N\}$ should be read modulo $2\pi$.

Unfortunately, it turns out that the $\mathcal{H}_r$-flows are not complete on the bounded phase space $\Sigma_g \times T$ [R3]. To remedy this incompleteness, it is needed to compactify the phase space in a suitable manner. For this purpose the key observation from [R3] is that it turns out possible to embed the noncomplete phase space $\Sigma_g \times T$ densely and symplectically in $\mathbb{CP}^N$. Here the complex projective space is to be viewed as a $2N$-dimensional real manifold with symplectic form proportional to the standard symplectic form inherited from the Fubini-Study Kähler metric on $\mathbb{CP}^N$. The Hamiltonians $\mathcal{H}_1, \ldots, \mathcal{H}_N$ (2.15) lift under this embedding to smooth (Poisson commuting) Hamiltonians on $\mathbb{CP}^N$ [R3] and the completeness of the corresponding Hamiltonian flows is thus immediate from the compactness of the extended phase space $\mathbb{CP}^N$.

The relevant embedding of the noncomplete phase space $\Sigma_g \times T$ into $\mathbb{CP}^N$ presented by Ruijsenaars is given explicitly by $(\mathbf{x}, \mathbf{p}) \mapsto [1 : z_1 : z_2 : \cdots : z_N] \in \mathbb{CP}^N$ with

$$(2.18) \qquad z_j = e^{i\langle \omega_j, \mathbf{p} \rangle} \left( \frac{\langle \mathbf{a}_j, \mathbf{x} \rangle - g}{2\pi/\alpha - g - \langle \mathbf{a}_{max}, \mathbf{x} \rangle} \right)^{1/2}, \qquad j = 1, \ldots, N.$$

The inverse mapping $[z_0 : z_1 : z_2 : \cdots : z_N] \mapsto (\mathbf{x}, \mathbf{p}) \in \Sigma_g \times T$, defined on an open dense patch $\{ [z_0 : \cdots : z_N] \mid z_j \neq 0 \ (j = 0, \ldots, N) \}$ of $\mathbb{CP}^N$, reads

$$(2.19a) \qquad \langle \mathbf{a}_j, \mathbf{x} \rangle = (2\pi/\alpha - (N+1)g) \frac{|z_j|^2}{|z_0|^2 + \cdots + |z_N|^2} + g,$$

$$(2.19b) \qquad e^{i\langle \omega_j, \mathbf{p} \rangle} = \frac{z_j |z_0|}{z_0 |z_j|}$$

$j = 1, \ldots, N$. (This gives the components of $\mathbf{x}, \mathbf{p}$ with respect to the bases $\{\mathbf{a}_1, \ldots, \mathbf{a}_N\}$ and $\{\omega_1, \ldots, \omega_N\}$, respectively.) The above mappings are symplectic when $\Sigma_g \times T$ is equipped with the standard symplectic form induced by $\sum_{j=1}^{N+1} dx_j \wedge dp_j$ and $\mathbb{CP}^N$ is endowed with the renormalized Fubini-Study symplectic form

$$(2.20) \quad \omega_R = \frac{2iR^2}{\sum_{j=0}^{N} |z_j|^2} \left( \sum_{j=1}^{N} dz_j \wedge d\bar{z}_j - \frac{1}{\sum_{j=0}^{N} |z_j|^2} \Big( \sum_{j=1}^{N} \bar{z}_j dz_j \Big) \wedge \Big( \sum_{j=1}^{N} z_j d\bar{z}_j \Big) \right),$$

where the normalization is such that the integral of $\omega_R$ over a complex projective line equals $4\pi R^2$ with

$$(2.21) \qquad 2R^2 = \frac{2\pi}{\alpha} - (N+1)g.$$

The coordinate functions in (2.19a) clearly extend to smooth functions on the whole of $\mathbb{CP}^N$. The image of the extension of the coordinate map to the completed phase space $\mathbb{CP}^N$ is therefore given by the compactification $\overline{\Sigma}_g$ of $\Sigma_g$ in $E$

$$(2.22) \qquad \overline{\Sigma}_g = \{ \mathbf{x} \in E \mid \langle \mathbf{a}_j, \mathbf{x} \rangle \geq g \ (j = 1, \ldots, N); \ \langle \mathbf{a}_{max}, \mathbf{x} \rangle \leq \frac{2\pi}{\alpha} - g \ \}.$$

It is natural to interpret the simplex $\overline{\Sigma}_g$ as the configuration space for the compactified trigonometric Ruijsenaars-Schneider model, even though globally the completed phase space $\mathbb{CP}^N$ has not a topology of product form. (In particular $\mathbb{CP}^N \not\cong \overline{\Sigma}_g \times T$.) Notice in this connection that the coordinate functions in (2.19b) for the momentum-like variables do *not* extend continuously to the boundary hyperplanes $z_j = 0$, $j = 0, \ldots, N$ (as the limiting value of (2.19b) for $z_j \to 0$ along a radius in the complex plane depends on the argument).

It is quite instructive to view how the compactification works topologically in the situation of two particles ($N = 1$). In this special case the reduced phase space $\Sigma_g \times T$ before completion has the structure of an open line segment ($\Sigma_g = \{(x/2, -x/2) \mid x \in\,]g, 2\pi/\alpha - g[\,\}$) times a real one-dimensional torus $\mathbb{T}^1$. Topologically this is a cylinder without the two boundary circles or, equivalently, a two-sphere with two distinct points extracted. The compactification adds the two extracted points (pinching) thus resulting in a compact phase space with the topology of a two-sphere $S^2$ ($\cong \mathbb{CP}^1$). The canonical projection $\Pi : \Sigma_g \times T \mapsto \Sigma_g$ clearly extends uniquely to a continuous projection $\overline{\Pi}$ of $S^2$ onto the closed line segment $\overline{\Sigma}_g = \{(x/2, -x/2) \mid x \in [g, 2\pi/\alpha - g]\,\})$, however, the fiber $\overline{\Pi}^{-1}(m)$ with $m \in \overline{\Sigma}_g$ reduces to a point when $m$ lies on the boundary $\overline{\Sigma}_g \setminus \Sigma_g$ whereas it is isomorphic to a real one-torus $\mathbb{T}^1$ for $m$ in the interior $\Sigma_g$. In particular, we do *not* have that $S^2$ is isomorphic to $\overline{\Sigma}_g \times \mathbb{T}^1$. See Figure 2.

FIGURE 2. The compactification of the phase space for $N = 1$ turning the cylinder $\Sigma_g \times T$ into a two-sphere ($\cong \mathbb{CP}^1$) by adding two points. The arrows indicate the canonical projections $\Pi, \overline{\Pi}$ onto the configurations spaces $\Sigma_g$ (open line segment) and $\overline{\Sigma}_g$ (closed line segment), respectively.

## 3. Quantization

In this section we quantize the compactified Ruijsenaars-Schneider model of the previous section. The quantum versions of the Hamiltonians $\mathcal{H}_1, \ldots, \mathcal{H}_N$ (2.15) for the relative particle motion in the center-of-mass frame will be given by commuting discrete difference operators acting in a finite-dimensional Hilbert space of functions with support on a finite uniform lattice over the classical compactified configuration space $\overline{\Sigma}_g$ (2.22).

### 3.1. Ruijsenaars difference operators.
In [R1] Ruijsenaars showed how the Poisson-commuting Hamiltonians $H_r$ (2.14) may be quantized formally (i.e., without specifying a Hilbert space) by means of canonical quantization in such a way that integrability is preserved. For the reduced integrals $\mathcal{H}_r$ (2.15) the procedure leads to difference operators of the form

$$\hat{\mathcal{H}}_r = (\hat{\mathcal{H}}_r^+ + \hat{\mathcal{H}}_r^-)/2, \qquad r = 1, \ldots, N, \tag{3.1}$$

with

$$\hat{\mathcal{H}}_r^+ = \sum_{\nu \in S_{N+1}(\omega_r)} V_\nu^{1/2}(\mathbf{x})\, T_\nu\, V_\nu^{1/2}(-\mathbf{x}), \tag{3.2a}$$

$$\hat{\mathcal{H}}_r^- = \sum_{\nu \in S_{N+1}(\omega_r)} V_\nu^{1/2}(-\mathbf{x})\, T_\nu^{-1}\, V_\nu^{1/2}(\mathbf{x}), \tag{3.2b}$$

where

$$T_\nu = e^{\langle \nu, \frac{\partial}{\partial \mathbf{x}} \rangle} \tag{3.3}$$

denotes the operator acting on functions $f: E \to \mathbb{C}$ by a translation over the weight vector $\nu$, i.e. $(T_\nu f)(\mathbf{x}) = f(\mathbf{x} + \nu)$, and the coefficients are determined by

$$V_\nu(\mathbf{x}) = \prod_{\substack{\mathbf{a} \in A_N \\ \langle \mathbf{a}, \nu \rangle = 1}} \frac{\sin \frac{\alpha}{2}(g + \langle \mathbf{a}, \mathbf{x} \rangle)}{\sin \frac{\alpha}{2} \langle \mathbf{a}, \mathbf{x} \rangle}. \tag{3.4}$$

The commutativity of the above difference operators is by no means evident from their explicit expressions, but it does follow immediately from Ruijsenaars' results in [R1]. To this end it is helpful to observe that $\hat{\mathcal{H}}_r^- = \hat{\mathcal{H}}_{N+1-r}^+$ because $-\omega_r$ is in the $S_{N+1}$-orbit of $\omega_{N+1-r}$ and $V_\nu(-\mathbf{x}) = V_{-\nu}(\mathbf{x})$. Hence, the commutativity already follows from the commutativity of $\hat{\mathcal{H}}_1^+, \ldots, \hat{\mathcal{H}}_N^+$, which can be traced back to [R1] by noticing that $\hat{\mathcal{H}}_r^+$ corresponds in Ruijsenaars' notation to the operator $\hat{S}_r\, \hat{S}_{N+1}^{-r/(N+1)}$ (with the number of particles of course being equal to $N+1$).

**Proposition 3.1** (Quantum integrability [R1]). *The difference operators $\hat{\mathcal{H}}_r^+$, $\hat{\mathcal{H}}_r^-$ and $\hat{\mathcal{H}}_r$, $r = 1, \ldots, N$ mutually commute.*

To see that the classical version of $\hat{\mathcal{H}}_r$ indeed amounts to $\mathcal{H}_r$ (2.15), one observes that after substituting $T_\nu = \exp(i\langle \nu, \mathbf{p} \rangle)$ (which is the classical analog of (3.3)) in $\hat{\mathcal{H}}_r$ one arrives at $\mathcal{H}_r$ (2.15) by using the identity

$$V_\nu(\mathbf{x}) V_\nu(-\mathbf{x}) = \prod_{\substack{\mathbf{a} \in A_N \\ \langle \mathbf{a}, \nu \rangle = 1}} \left( 1 - \frac{\sin^2(\frac{\alpha g}{2})}{\sin^2 \frac{\alpha}{2} \langle \mathbf{a}, \mathbf{x} \rangle} \right).$$

3.2. **The finite-dimensional Hilbert space.** The formal difference operators $\hat{\mathcal{H}}_r$ (3.1) and $\hat{\mathcal{H}}_r^\pm$ (3.2a), (3.2b) shift function arguments over vectors in the weight lattice $\Lambda$ (2.5). We will now assign a precise meaning to these difference operators as operators in a Hilbert space of functions with support in a uniform lattice over the classical compactified configuration space $\overline{\Sigma}_g$ (2.22).

The point $\rho$ (2.13) denotes the "minimal" vertex of the simplex $\overline{\Sigma}_g$ determined (uniquely) by the property that the functionals $\langle \mathbf{a}_j, \cdot \rangle$, $j = 1, \ldots, N$ simultaneously assume their minimum value $g$. By shifting from $\rho$ over vectors in the weight lattice $\Lambda$ (2.5), one generates a uniform lattice in $\overline{\Sigma}_g$ consisting of the points $\rho + \mu$, $\mu \in \Lambda^+$ (2.7) with $\langle \mathbf{a}_{max}, \rho + \mu \rangle = N g + \langle \mathbf{a}_{max}, \mu \rangle \leq \frac{2\pi}{\alpha} - g$. When the (positive) coupling constant $g$ and scale factor $\alpha$ are related by

$$\text{(3.5)} \qquad \frac{2\pi}{\alpha} - (N+1)g = M \in \mathbb{N} \setminus \{0\},$$

then the maximum value $\frac{2\pi}{\alpha} - g$ of the functional $\langle \mathbf{a}_{\max}, \cdot \rangle$ is assumed on the lattice. More to the point, it means that in this situation apart from the "minimal vertex" $\rho$ also the $N$ other vertices of the simplex $\overline{\Sigma}_g$ (2.22) (viz. the "maximal vertices" $\rho + M\omega_r$, $r = 1, \ldots, N$) lie on the lattice and, hence, that the lattice fits precisely over the classical configuration space including its boundary. See Figure 3.

FIGURE 3. The lattice $\rho + \Lambda_M^+$ over the configuration space $\overline{\Sigma}_g$ supporting the quantum wave functions for $N = 2$ and $M = 5$. The dimension of the Hilbert space $L^2(\rho + \Lambda_M^+)$ (i.e. the number of points in the lattice) amounts in this case to $\binom{2+5}{2} = 21$.

From now on we will assume that the condition in (3.5) is satisfied. (Notice that the condition is compatible with the parameter restriction in (2.16).) Let $\Lambda_M^+$ be the alcove of dominant weights in $\Lambda^+$ (2.7) given by

$$\text{(3.6)} \qquad \Lambda_M^+ = \{\lambda \in \Lambda^+ \mid \langle \mathbf{a}_{max}, \lambda \rangle \leq M\}$$

and let $L^2(\rho + \Lambda_M^+)$ be the finite-dimensional Hilbert space of complex functions over the lattice $\rho + \Lambda_M^+ := \{\rho + \mu \mid \mu \in \Lambda_M^+\}$ endowed with the (standard) sesquilinear inner product

$$\text{(3.7)} \qquad (f, h) = \sum_{\mu \in \Lambda_M^+} f(\rho + \mu)\overline{h(\rho + \mu)}$$

($f, h \in L^2(\rho + \Lambda_M^+)$). Notice that our conventions are such that the inner product is linear in the first slot and antilinear in the second slot. The dimension of the Hilbert space $L^2(\rho + \Lambda_M^+)$ amounts to the number of points in $\Lambda_M^+$, i.e., the number of vectors of the form $n_1\omega_1 + \cdots + n_N\omega_N$ with $n_j \in \mathbb{N}$ $(j = 1, \ldots, N)$ and $\langle \mathbf{a}_{max}, n_1\omega_1 + \cdots + n_N\omega_N \rangle = n_1 + \cdots + n_N \leq M$:

$$\dim (L^2(\rho + \Lambda_M^+)) = \frac{(N+M)!}{N!\, M!}. \tag{3.8}$$

In order to see that the difference operators $\hat{\mathcal{H}}_r^{(\pm)}$ (3.1), (3.2a), (3.2b) are well-defined as operators in $L^2(\rho + \Lambda_M^+)$ we need the following lemma.

**Lemma 3.2** (Regularity, positivity and vanishing boundary conditions). *For positive parameters $\alpha, g$ subject to the condition (3.5), $\mu \in \Lambda_M^+$ (3.6) and $\nu$ in the $S_{N+1}$-orbit of a fundamental weight vector $\omega_r$ (2.3), one has that*

$$0 < V_\nu(\rho + \mu) < \infty, \ 0 < V_\nu(-\rho - \mu - \nu) < \infty \ \ if \ \ \mu + \nu \in \Lambda_M^+,$$
$$V_\nu(\rho + \mu) = 0 \ \ if \ \ \mu + \nu \notin \Lambda_M^+$$

*and, if in addition $g \notin \{\frac{1}{N}, \frac{1}{N-1}, \frac{1}{N-2}, \cdots, \frac{1}{2}, 1\}$, that*

$$-\infty < V_\nu(-\rho - \mu - \nu) < \infty \ \ if \ \ \mu + \nu \notin \Lambda_M^+$$

*(where $\rho$ and $V_\nu(\mathbf{x})$ are given by (2.13) and (3.4), respectively).*

*Proof.* Let us write

$$V_\nu(\rho + \mu) = \prod_{\substack{\mathbf{a} \in A_N^+ \\ \langle \mathbf{a}, \nu \rangle = 1}} \frac{\sin \frac{\alpha}{2}(\langle \mathbf{a}, \rho + \mu \rangle + g)}{\sin \frac{\alpha}{2}\langle \mathbf{a}, \rho + \mu \rangle} \prod_{\substack{\mathbf{a} \in A_N^+ \\ \langle \mathbf{a}, \nu \rangle = -1}} \frac{\sin \frac{\alpha}{2}(\langle \mathbf{a}, \rho + \mu \rangle - g)}{\sin \frac{\alpha}{2}\langle \mathbf{a}, \rho + \mu \rangle}.$$

From the inequality

$$0 < g \leq \langle \mathbf{a}, \rho + \mu \rangle \leq \frac{2\pi}{\alpha} - g < \frac{2\pi}{\alpha} \tag{$*$}$$

for $\mathbf{a} \in A_N^+$ and $\mu \in \Lambda_M^+$, it is seen that all factors in the denominator of the above formula for $V_\nu(\rho + \mu)$ are positive and that all factors in the numerator are nonnegative. Zeros in the numerator appear when $\langle \mathbf{a}, \rho + \mu \rangle + g$ becomes equal to $2\pi/\alpha$ or when $\langle \mathbf{a}, \rho + \mu \rangle - g$ becomes equal to 0. The first situation occurs if and only if $\langle \mathbf{a}_{max}, \nu \rangle = 1$ and $\langle \mathbf{a}_{max}, \mu \rangle = M$, i.e., iff $\langle \mathbf{a}_{max}, \mu + \nu \rangle > M$. The second situation occurs if and only if $\langle \mathbf{a}_j, \nu \rangle = -1$ and $\langle \mathbf{a}_j, \mu \rangle = 0$ for certain simple root $\mathbf{a}_j$ (2.2), i.e., iff $\langle \mathbf{a}_j, \mu + \nu \rangle < 0$ for certain simple root $\mathbf{a}_j$.

In a similar way one derives from the formula

$$V_\nu(-\rho - \mu - \nu) =$$
$$\prod_{\substack{\mathbf{a} \in A_N^+ \\ \langle \mathbf{a}, \nu \rangle = 1}} \frac{\sin \frac{\alpha}{2}(\langle \mathbf{a}, \rho + \mu \rangle + 1 - g)}{\sin \frac{\alpha}{2}(\langle \mathbf{a}, \rho + \mu \rangle + 1)} \prod_{\substack{\mathbf{a} \in A_N^+ \\ \langle \mathbf{a}, \nu \rangle = -1}} \frac{\sin \frac{\alpha}{2}(\langle \mathbf{a}, \rho + \mu \rangle - 1 + g)}{\sin \frac{\alpha}{2}(\langle \mathbf{a}, \rho + \mu \rangle - 1)}$$

combined with the inequality $(*)$, that $V_\nu(-\rho - \mu - \nu)$ is positive and finite for $\mu \in \Lambda_M^+$ with $\mu + \nu \in \Lambda_M^+$. The denominators become zero when $\langle \mathbf{a}, \rho + \mu \rangle + 1 = 2\pi/\alpha$ or when $\langle \mathbf{a}, \rho + \mu \rangle - 1 = 0$, which can happen only if $\mu + \nu \notin \Lambda_M^+$ and $g \in \{\frac{1}{N}, \frac{1}{N-1}, \frac{1}{N-2}, \cdots, \frac{1}{2}, 1\}$. □

We learn from Lemma 3.2 that for parameters subject to (3.5) the coefficient functions $V_\nu(\mathbf{x})$ (3.4) are regular and positive on the lattice points $\rho+\mu$, $\mu \in \Lambda_M^+$ with $\mu+\nu \in \Lambda_M^+$ and zero on the boundary lattice points $\rho+\mu$, $\mu \in \Lambda_M^+$ with $\mu+\nu \notin \Lambda_M^+$. Similarly, the coefficient function $V_\nu(-\mathbf{x}-\nu)$ is regular and positive on the lattice points $\rho+\mu$ with $\mu, \mu+\nu \in \Lambda_M^+$ and generically (i.e. for $g \neq 1/j$, $j = 1, \ldots, N$) regular on the boundary lattice points $\rho+\mu$ with $\mu \in \Lambda_M^+$, $\mu+\nu \notin \Lambda_M^+$. (Shifts of the type $\mp\mathbf{x} \to \mp\mathbf{x} - \nu$ in the functions governing the coefficients of $\hat{\mathcal{H}}_r^\pm$ (3.2a), (3.2b) originate from commuting the coefficients on the right of the translation operator $T_\nu^{\pm 1}$ to the left.) This entails that for parameters subject to (3.5) (and $g \notin \{\frac{1}{N}, \frac{1}{N-1}, \ldots, 1\}$) the difference operators $\hat{\mathcal{H}}_1^\pm, \ldots, \hat{\mathcal{H}}_N^\pm$ (3.2a), (3.2b) (and hence $\hat{\mathcal{H}}_1, \ldots, \hat{\mathcal{H}}_N$ (3.1)) admit a well-defined restriction to functions in $L^2(\rho+\Lambda_M^+)$ which maps the Hilbert space into itself. The explicit action on functions $f \in L^2(\rho+\Lambda_M^+)$ is given by

$$(3.9) \qquad (\hat{\mathcal{H}}_r^\pm f)(\rho+\mu) = \sum_{\nu \in S_{N+1}(\omega_r)} W_\nu^\pm(\rho+\mu) f(\rho+\mu+\nu)$$

($\mu \in \Lambda_M^+$) with

$$W_\nu^+(\rho+\mu) = \begin{cases} V_\nu^{1/2}(\rho+\mu) V_\nu^{1/2}(-\rho-\mu-\nu) > 0 & for \quad \mu+\nu \in \Lambda_M^+ \\ 0 & for \quad \mu+\nu \notin \Lambda_M^+, \end{cases}$$

$$W_\nu^-(\rho+\mu) = \begin{cases} V_\nu^{1/2}(-\rho-\mu) V_\nu^{1/2}(\rho+\mu-\nu) > 0 & for \quad \mu-\nu \in \Lambda_M^+ \\ 0 & for \quad \mu-\nu \notin \Lambda_M^+ \end{cases}$$

(recall to this end also that $V_\nu(-\mathbf{x}) = V_{-\nu}(\mathbf{x})$). The vanishing boundary conditions for the coefficients $W_\nu^\pm(\rho+\mu)$ guarantee that $(\hat{\mathcal{H}}_r^\pm f)(\rho+\mu)$, $\mu \in \Lambda_M^+$ depends only on the values of $f(\cdot)$ in the points of the lattice $\rho+\Lambda_M^+$, i.e., we have that $\hat{\mathcal{H}}_r^\pm$ is well-defined as operator in $L^2(\rho+\Lambda_M^+)$. For $g \in \{\frac{1}{N}, \frac{1}{N-1}, \ldots, 1\}$ ambiguities in the value of the products $V_\nu(\rho+\mu) V_\nu(-\rho-\mu-\nu)$ and $V_\nu(-\rho-\mu) V_\nu(\rho+\mu-\nu)$ at the boundary points $\mu \in \Lambda_M^+$ with $\mu+\nu \notin \Lambda_M^+$ may arise as zeros deriving from $V_\nu(\rho+\mu)$ and $V_\nu(-\rho-\mu)$ can meet with possible singularities deriving from $V_\nu(-\rho-\mu-\nu)$ and $V_\nu(\rho+\mu-\nu)$, respectively. For instance, when $g = 1$ we have for generic $\mathbf{x} \in \mathbb{R}^N$ that $V_\nu(\mathbf{x}) V_\nu(-\mathbf{x}-\nu) \equiv 1$. Here we will resolve such ambiguities in the value of coefficients at the boundary lattice points for $g$ in the exceptional set $\{\frac{1}{N}, \frac{1}{N-1}, \ldots, 1\}$ by requiring continuity under small variations in $g$. Specifically, this means that for all parameter values subject to the condition (3.5) we will pick the action of $\hat{\mathcal{H}}_r^\pm$ with vanishing boundary conditions in accordance with (3.9).

**Proposition 3.3** (Self-adjointness). *Let us assume (positive) parameters subject to the condition (3.5) and let the action of $\hat{\mathcal{H}}_r^\pm : L^2(\rho+\Lambda_M^+) \to L^2(\rho+\Lambda_M^+)$ be given by (3.9). Then the difference operators $\hat{\mathcal{H}}_r = (\hat{\mathcal{H}}_r^+ + \hat{\mathcal{H}}_r^-)/2$, $r = 1, \ldots, N$ are self-adjoint in $L^2(\rho+\Lambda_M^+)$.*

*Proof.* It suffices to demonstrate that the operators $\hat{\mathcal{H}}_r^+$ and $\hat{\mathcal{H}}_r^-$ are each others' adjoints in $L^2(\rho + \Lambda_M^+)$. Some elementary manipulations produce:

$$
\begin{aligned}
(\hat{\mathcal{H}}_r^+ f, h) &= \sum_{\mu \in \Lambda_M^+} (\hat{\mathcal{H}}_r^+ f)(\rho+\mu)\overline{h(\rho+\mu)} \\
&\stackrel{(3.9)}{=} \sum_{\nu \in S_{N+1}(\omega_r)} \sum_{\mu \in \Lambda_M^+} W_\nu^+(\rho+\mu) f(\rho+\mu+\nu)\overline{h(\rho+\mu)} \\
&\stackrel{(i)}{=} \sum_{\nu \in S_{N+1}(\omega_r)} \sum_{\substack{\mu \in \Lambda_M^+ \\ \mu+\nu \in \Lambda_M^+}} W_\nu^+(\rho+\mu) f(\rho+\mu+\nu)\overline{h(\rho+\mu)} \\
&\stackrel{(ii)}{=} \sum_{\nu \in S_{N+1}(\omega_r)} \sum_{\substack{\tilde{\mu} \in \Lambda_M^+ \\ \tilde{\mu}-\nu \in \Lambda_M^+}} W_\nu^+(\rho+\tilde{\mu}-\nu) f(\rho+\tilde{\mu})\overline{h(\rho+\tilde{\mu}-\nu)} \\
&\stackrel{(iii)}{=} \sum_{\nu \in S_{N+1}(\omega_r)} \sum_{\substack{\tilde{\mu} \in \Lambda_M^+ \\ \tilde{\mu}-\nu \in \Lambda_M^+}} f(\rho+\tilde{\mu})\overline{W_\nu^-(\rho+\tilde{\mu}) h(\rho+\tilde{\mu}-\nu)} \\
&\stackrel{(i),(3.9)}{=} \sum_{\tilde{\mu} \in \Lambda_M^+} f(\rho+\tilde{\mu})\overline{(\hat{\mathcal{H}}_r^- h)(\rho+\tilde{\mu})} = (f, \hat{\mathcal{H}}_r^- h),
\end{aligned}
$$

where we have used (i) the vanishing boundary conditions for the coefficients $W_\nu^\pm$, (ii) the substitution $\mu = \tilde{\mu} - \nu$ and (iii) that $W_\nu^+(\rho + \tilde{\mu} - \nu) = W_\nu^-(\rho + \tilde{\mu}) \geq 0$. □

*Remark:* For given trigonometric period $\frac{2\pi}{\alpha} > 1$, the parameter restriction in (3.5) determines a quantization condition on the coupling parameter $g$ (permitting only a finite number of values for $g$ labeled by $M \in \{1, \ldots, [\frac{2\pi}{\alpha}]\}$). However, it is also possible (and probably somewhat more natural) to instead interpret the restriction in (3.5) as a quantization condition on a step size parameter. Recall to this end that in the present paper we have scaled our variables such that the scale parameter $\beta$ appearing in the classical (compactified) Ruijsenaars-Schneider Hamiltonian $H$ (1.1) has the value 1 (see the note at the end of the introduction). By substituting $x_j \to \beta^{-1} x_j$ and $\alpha \to \alpha\beta$ ($\beta > 0$) in the difference operators of Sect. 3.1, we reintroduce the scale parameter $\beta$ in our quantum Hamiltonians. Specifically, the operators $\hat{\mathcal{H}}_r^{(\pm)}$ (3.1), (3.2a), (3.2b) then pass over to discrete difference operators of the form given in (3.1)-(3.4) with the coupling parameter $g$ and the translation operator $T_\nu$ (3.3) replaced by $\beta g$ and $T_\nu = \exp(\beta \langle \nu, \frac{\partial}{\partial \mathbf{x}} \rangle)$, respectively. In other words, at the quantum level the scale parameter $\beta$ enters as the step size parameter of the discrete difference operators [R1]. As a consequence of this rescaling, the lattice supporting the wave functions is going to be scaled by $\beta$ resulting in a lattice of the form $\beta(\rho+\Lambda_M^+)$ and the parameter restriction in (3.5) passes over to the condition $\frac{2\pi}{\alpha\beta} - (N+1)g = M \in \mathbb{N}\setminus\{0\}$. For a given trigonometric period $\frac{2\pi}{\alpha}$ and (positive) coupling parameter $g$, the latter parameter restriction may be interpreted as a quantization condition on the step size parameter $\beta$ (permitting an infinite series of values for $\beta$ labeled by $M \in \mathbb{N}\setminus\{0\}$). The quantization condition at issue adjusts the step size such that the lattice $\beta(\rho+\Lambda_M^+)$ fits precisely over the classical configuration space $\overline{\Sigma}_{\beta g}$ (cf. (2.22)) including the corner points (vertices).

## 4. Wave functions

In this section an orthonormal basis for $L^2(\rho+\Lambda_M^+)$ is presented consisting of joint eigenfunctions of $\hat{\mathcal{H}}_1, \ldots, \hat{\mathcal{H}}_N$.

### 4.1. A factorized eigenfunction. Let

$$(4.1) \qquad \Delta(\mu) = \frac{1}{C_+(\mu)\, C_-(\mu)}, \qquad \mu \in \Lambda^+$$

with

$$(4.2a) \qquad C_+(\mu) = \prod_{\mathbf{a}\in A_N^+} \frac{(\langle \mathbf{a},\rho\rangle : \sin_\alpha)_{\langle \mathbf{a},\mu\rangle}}{(g+\langle \mathbf{a},\rho\rangle : \sin_\alpha)_{\langle \mathbf{a},\mu\rangle}},$$

$$(4.2b) \qquad C_-(\mu) = \prod_{\mathbf{a}\in A_N^+} \frac{(1-g+\langle \mathbf{a},\rho\rangle : \sin_\alpha)_{\langle \mathbf{a},\mu\rangle}}{(1+\langle \mathbf{a},\rho\rangle : \sin_\alpha)_{\langle \mathbf{a},\mu\rangle}}.$$

Here we have introduced "trigonometric Pochhammer symbols" defined by

$$(4.3) \qquad (z : \sin_\alpha)_m := \begin{cases} 1 & m=0 \\ \prod_{k=0}^{m-1} \sin\frac{\alpha}{2}(z+k) & m=1,2,3,\ldots \end{cases}$$

We need two preparatory lemmas. The first states that, for positive parameters subject to (3.5), the value of $C_\pm(\mu)$ (and hence that of $\Delta(\mu)$) is positive and finite for $\mu \in \Lambda_M^+$ (3.6); the second lemma describes a functional relation between $\Delta$ (4.1) and the coefficient functions $V_\nu(\mathbf{x})$ (3.4), $\nu \in S_{N+1}(\omega_r)$.

**Lemma 4.1** (Regularity and positivity). *For positive parameters $\alpha$, $g$ subject to the condition (3.5), one has that*

$$C_\pm(\mu) > 0 \quad for \quad \mu \in \Lambda_M^+$$

*(with $\Lambda_M^+$ given by (3.6)).*

*Proof.* Using inequality (∗) in the proof of Lemma 3.2, it is not difficult to infer that the arguments of the sine factors in $C_\pm(\mu)$ (4.2a), (4.2b) lie between 0 and $\pi/2$. □

**Lemma 4.2.** *Let $\nu$ be in the $S_{N+1}$-orbit of a fundamental weight vector $\omega_r$ (2.3) and let $\mu, \mu+\nu \in \Lambda^+$ (2.7). Then*

$$\Delta(\mu+\nu)V_\nu(-\rho-\mu-\nu) = \Delta(\mu)V_\nu(\rho+\mu).$$

*Proof.* We have that

$$\Delta(\mu+\nu) = \prod_{\mathbf{a}\in A_N^+} \frac{(g+\langle\mathbf{a},\rho\rangle:\sin_\alpha)_{\langle\mathbf{a},\mu+\nu\rangle}}{(1-g+\langle\mathbf{a},\rho\rangle:\sin_\alpha)_{\langle\mathbf{a},\mu+\nu\rangle}} \frac{(1+\langle\mathbf{a},\rho\rangle:\sin_\alpha)_{\langle\mathbf{a},\mu+\nu\rangle}}{(\langle\mathbf{a},\rho\rangle:\sin_\alpha)_{\langle\mathbf{a},\mu+\nu\rangle}}$$

$$= \prod_{\mathbf{a}\in A_N^+} \frac{(g+\langle\mathbf{a},\rho\rangle:\sin_\alpha)_{\langle\mathbf{a},\mu\rangle}}{(1-g+\langle\mathbf{a},\rho\rangle:\sin_\alpha)_{\langle\mathbf{a},\mu\rangle}} \frac{(1+\langle\mathbf{a},\rho\rangle:\sin_\alpha)_{\langle\mathbf{a},\mu\rangle}}{(\langle\mathbf{a},\rho\rangle:\sin_\alpha)_{\langle\mathbf{a},\mu\rangle}}$$

$$\times \prod_{\substack{\mathbf{a}\in A_N^+ \\ \langle\mathbf{a},\nu\rangle=1}} \frac{\sin\frac{\alpha}{2}(\langle\mathbf{a},\rho+\mu\rangle+g)}{\sin\frac{\alpha}{2}(\langle\mathbf{a},\rho+\mu\rangle+1-g)} \frac{\sin\frac{\alpha}{2}(\langle\mathbf{a},\rho+\mu\rangle+1)}{\sin\frac{\alpha}{2}\langle\mathbf{a},\rho+\mu\rangle}$$

$$\times \prod_{\substack{\mathbf{a}\in A_N^+ \\ \langle\mathbf{a},\nu\rangle=-1}} \frac{\sin\frac{\alpha}{2}(\langle\mathbf{a},\rho+\mu\rangle-g)}{\sin\frac{\alpha}{2}(\langle\mathbf{a},\rho+\mu\rangle-1+g)} \frac{\sin\frac{\alpha}{2}(\langle\mathbf{a},\rho+\mu\rangle-1)}{\sin\frac{\alpha}{2}\langle\mathbf{a},\rho+\mu\rangle}.$$

Multiplication by

$$V_\nu(-\rho-\mu-\nu) = \prod_{\substack{\mathbf{a}\in A_N^+ \\ \langle\mathbf{a},\nu\rangle=1}} \frac{\sin\frac{\alpha}{2}(\langle\mathbf{a},\rho+\mu\rangle+1-g)}{\sin\frac{\alpha}{2}(\langle\mathbf{a},\rho+\mu\rangle+1)} \prod_{\substack{\mathbf{a}\in A_N^+ \\ \langle\mathbf{a},\nu\rangle=-1}} \frac{\sin\frac{\alpha}{2}(\langle\mathbf{a},\rho+\mu\rangle-1+g)}{\sin\frac{\alpha}{2}(\langle\mathbf{a},\rho+\mu\rangle-1)}$$

leads after cancellation of common terms in the numerator and the denominator to an expression of the form

$$\prod_{\mathbf{a}\in A_N^+} \frac{(g+\langle\mathbf{a},\rho\rangle:\sin_\alpha)_{\langle\mathbf{a},\mu\rangle}}{(1-g+\langle\mathbf{a},\rho\rangle:\sin_\alpha)_{\langle\mathbf{a},\mu\rangle}} \frac{(1+\langle\mathbf{a},\rho\rangle:\sin_\alpha)_{\langle\mathbf{a},\mu\rangle}}{(\langle\mathbf{a},\rho\rangle:\sin_\alpha)_{\langle\mathbf{a},\mu\rangle}}$$

$$\times \prod_{\substack{\mathbf{a}\in A_N^+ \\ \langle\mathbf{a},\nu\rangle=1}} \frac{\sin\frac{\alpha}{2}(\langle\mathbf{a},\rho+\mu\rangle+g)}{\sin\frac{\alpha}{2}\langle\mathbf{a},\rho+\mu\rangle} \prod_{\substack{\mathbf{a}\in A_N^+ \\ \langle\mathbf{a},\nu\rangle=-1}} \frac{\sin\frac{\alpha}{2}(\langle\mathbf{a},\rho+\mu\rangle-g)}{\sin\frac{\alpha}{2}\langle\mathbf{a},\rho+\mu\rangle}$$

$$= \Delta(\mu)V_\nu(\rho+\mu).$$

□

After these preliminaries we are now in the position to introduce a factorized joint eigenfunction of $\hat{\mathcal{H}}_1,\ldots,\hat{\mathcal{H}}_N$. Let $\Psi_\mathbf{0} : (\rho+\Lambda_M^+) \to \mathbb{R}$ be the lattice function defined by

(4.4) $$\Psi_\mathbf{0}(\rho+\mu) = \frac{1}{\mathcal{N}_0^{1/2}}\Delta^{1/2}(\mu), \qquad \mu \in \Lambda_M^+,$$

where the normalization constant $\mathcal{N}_0$ is chosen such that $(\Psi_\mathbf{0},\Psi_\mathbf{0}) = 1$ (recall the inner product (3.7)). Notice that $\Psi_\mathbf{0}$ is well-defined and positive at the lattice points $\rho+\mu$, $\mu \in \Lambda_M^+$ (3.6) because of Lemma 4.1.

**Proposition 4.3** (Factorized eigenfunction). *For positive parameters $\alpha$, $g$ subject to condition (3.5), the function $\Psi_{\mathbf{0}}$ (4.4) is a joint eigenfunction of the difference operators $\hat{\mathcal{H}}_r : L^2(\rho + \Lambda_M^+) \to L^2(\rho + \Lambda_M^+)$ (3.1), (3.9)*

$$\hat{\mathcal{H}}_r \Psi_{\mathbf{0}} = \Big( \sum_{\nu \in S_{N+1}(\omega_r)} \cos \alpha \langle \nu, \rho \rangle \Big) \Psi_{\mathbf{0}}, \qquad r = 1, \ldots, N$$

*(where $\rho$ is given by (2.13)).*

*Proof.* One has that

$$(\hat{\mathcal{H}}_r \Psi_{\mathbf{0}})(\rho + \mu) \stackrel{(3.1),(3.9)}{=} \frac{1}{2\mathcal{N}_0^{1/2}} \sum_{\nu \in S_{N+1}(\omega_r)} \Big( W_\nu^+(\rho + \mu) \Delta^{1/2}(\rho + \mu + \nu)$$
$$+ W_\nu^-(\rho + \mu) \Delta^{1/2}(\rho + \mu - \nu) \Big)$$
$$\stackrel{(i)}{=} \frac{1}{2\mathcal{N}_0^{1/2}} \Big( \sum_{\nu \in S_{N+1}(\omega_r)} V_\nu(\rho + \mu) + V_\nu(-\rho - \mu) \Big) \Delta^{1/2}(\rho + \mu)$$
$$\stackrel{(ii)}{=} \Big( \sum_{\nu \in S_{N+1}(\omega_r)} \cos \alpha \langle \nu, \rho \rangle \Big) \Psi_{\mathbf{0}}(\rho + \mu),$$

where we have used: $(i)$ Lemma 4.2 combined with the fact that $W_\nu^\pm(\rho + \mu) = V_{\pm\nu}(\rho + \mu) = 0$ for $\mu \pm \nu \notin \Lambda_M^+$ and $(ii)$ the Macdonald identity [M1, Theorem (2.8)] (see also Appendix B)

$$\sum_{\nu \in S_{N+1}(\omega_r)} V_\nu(\mathbf{x}) = \sum_{\nu \in S_{N+1}(\omega_r)} \cos \alpha \langle \nu, \rho \rangle.$$

□

The following proposition gives a compact product formula for the proportionality constant $\mathcal{N}_0$, which normalizes the wave function (4.4) such that its $L^2$-norm is equal to 1.

**Proposition 4.4** (Normalization). *The value of $\mathcal{N}_0$ is given by*

$$\mathcal{N}_0 = \sum_{\mu \in \Lambda_M^+} \Delta(\mu) = 2^{N(M-1)}(N+1) \prod_{1 \leq n \leq N} (1 + ng : \sin_\alpha)_{M-1},$$

*where it is assumed that the parameters satisfy condition (3.5).*

*Proof.* Clearly $\mathcal{N}_0 = \sum_{\mu \in \Lambda_M^+} \Delta(\mu)$ normalizes $\Psi_{\mathbf{0}}$ (4.4) such that $(\Psi_{\mathbf{0}}, \Psi_{\mathbf{0}}) = 1$. The evaluation of the sum leading to the product formula on the r.h.s. hinges on a terminating version of a recent summation formula due to Aomoto, Ito, and Macdonald [Ao, I, M5]. The details of the summation are relegated to Appendix A (see (A.4)). □

By specializing the Macdonald identity in the last line of the proof of Proposition 4.3 to $\mathbf{x} = \rho$ and recalling Lemma 3.2, one arrives at a simple product formula for the

eigenvalues:

$$(4.5) \quad \sum_{\nu \in S_{N+1}(\omega_r)} \cos\alpha\langle\nu,\rho\rangle = \sum_{\nu \in S_{N+1}(\omega_r)} V_\nu(\rho) \stackrel{\text{Lemma 3.2}}{=} V_{\omega_r}(\rho)$$

$$= \frac{\prod_{j=1}^{N+1} \sin(\frac{j\alpha g}{2})}{\prod_{j=1}^{r} \sin(\frac{j\alpha g}{2}) \prod_{j=1}^{N+1-r} \sin(\frac{j\alpha g}{2})}.$$

For $r=1$ this product formula specializes to the well-known geometric progression

$$(4.6) \quad \sum_{j=1}^{N+1} \cos(\alpha \rho_j) = \frac{\sin(\frac{\alpha g}{2}(N+1))}{\sin(\frac{\alpha g}{2})}.$$

4.2. **The complete eigenbasis.** We will now extend the factorized wave function $\Psi_\mathbf{0}$ (4.4) to an orthonormal basis of $L^2(\rho + \Lambda_M^+)$ consisting of joint eigenfunctions of the commuting operators $\hat{\mathcal{H}}_1, \ldots, \hat{\mathcal{H}}_N$. The eigenbasis will be expressed in terms of Macdonald polynomials with $|q|=1$. To describe these we need notation for the *elementary symmetric functions*

$$(4.7) \quad E_r^\pm(\mathbf{x}) = \sum_{\nu \in S_{N+1}(\omega_r)} e^{\pm i\alpha\langle\nu,\mathbf{x}\rangle}, \qquad r=1,\ldots,N$$

together with their real parts

$$(4.8) \quad E_r(\mathbf{x}) = \sum_{\nu \in S_{N+1}(\omega_r)} \cos\alpha\langle\nu,\mathbf{x}\rangle, \qquad r=1,\ldots,N$$

and also for the *monomial symmetric functions*

$$(4.9) \quad m_\lambda(\mathbf{x}) = \sum_{\mu \in S_{N+1}(\lambda)} e^{i\alpha\langle\mu,\mathbf{x}\rangle}, \qquad \lambda \in \Lambda^+.$$

Notice that $E_r^-(\mathbf{x}) = \overline{E_r^+(\mathbf{x})} = E_{N+1-r}^+(\mathbf{x})$ (since $-\omega_r$ lies in the $S_{N+1}$-orbit of $\omega_{N+1-r}$). The *Macdonald polynomials* $p_\lambda(\mathbf{x})$, $\lambda \in \Lambda^+$ are now defined as the unique trigonometric polynomials of the form

$$(4.10a) \quad p_\lambda(\mathbf{x}) = m_\lambda(\mathbf{x}) + \sum_{\substack{\mu \in \Lambda^+ \\ \mu \prec \lambda}} c_{\lambda,\mu} m_\mu(\mathbf{x}), \qquad c_{\lambda,\mu} \in \mathbb{R}$$

satisfying the difference equations

$$(4.10b) \quad \sum_{\nu \in S_{N+1}(\omega_r)} V_\nu(\mathbf{x}) p_\lambda(\mathbf{x}+\nu) = E_r^+(\rho+\lambda)\, p_\lambda(\mathbf{x}), \qquad r=1,\ldots,N$$

(with the coefficients $V_\nu(\mathbf{x})$ given by (3.4)). For generic parameters the existence of polynomials $p_\lambda$ of this form follows from the work of Macdonald in [M2, M3, M4] (see Appendix B for a brief summary of the results most relevant to us here). Our parameters $\alpha$ and $g$ are related to the parameters $q$ and $t$ employed by Macdonald via $t=q^g$ and $q=e^{i\alpha}$ (cf. Appendix B). So, in particular, we have that $|q|=1$ (and $|t|=1$) for real $\alpha$ (and $g$).

An important property of the Macdonald polynomials is that after renormalizing in the following way

$$(4.11) \qquad P_\lambda(\mathbf{x}) = C_+(\lambda)\, p_\lambda(\mathbf{x}), \qquad \lambda \in \Lambda^+$$

(where $C_+(\lambda)$ is given by (4.2a)), they satisfy the symmetry relations [Ko1, M4, EK] (cf. also Appendix B)

$$(4.12) \qquad P_\lambda(\rho + \mu) = P_\mu(\rho + \lambda), \qquad \lambda, \mu \in \Lambda^+.$$

The polynomials in (4.11) are normalized such that $P_\lambda(\rho) = 1$, as is clear from the symmetry relation (4.12) specialized to $\mu = \mathbf{0}$ (since $P_\mathbf{0}(\cdot) \equiv 1$).

Even though for generic parameters the existence of the Macdonald polynomials of the form (4.10a), (4.10b) is guaranteed by Macdonald's work, it is a priori not entirely obvious that it is possible to specialize them to positive parameter values for $\alpha, g$ subject to the constraint in (3.5). The point is that for certain special values of the parameters the eigenvalues $E_r^+(\rho + \lambda)$ on the r.h.s. of (4.10b) may not be semisimple. This manifests itself through possible singularities in the expansion coefficients $c_{\lambda,\mu}$ of (4.10a) at such special parameter values. The next lemma ensures that for $\lambda \in \Lambda_M^+$ the eigenvalue $E_r^+(\rho+\lambda)$ is in fact semisimple for positive parameters $\alpha, g$ subject to condition (3.5) and, hence, that the Macdonald polynomials $p_\lambda(\cdot), \lambda \in \Lambda_M^+$ indeed admit a well-defined specialization to these parameters values (without any singularities in the expansion coefficients being hit).

**Lemma 4.5** (Semisimple spectrum). *Let $\alpha, g$ be positive and subject to the condition (3.5). Then the elementary symmetric functions $E_1^+(\mathbf{x}), \ldots, E_N^+(\mathbf{x})$ (4.8) separate the points of the lattice $\rho + \Lambda_M^+$.*

*Proof.* For $\mathbf{x}, \mathbf{y} \in E$ (2.1), we have that $E_r^+(\mathbf{x}) = E_r^+(\mathbf{y})$ for $r = 1, \ldots, N$ if and only if

$$\mathbf{x} = \sigma(\mathbf{y}) \mod \frac{2\pi}{\alpha} Q,$$

with $\sigma \in S_{N+1}$. If both $\mathbf{x}$ and $\mathbf{y}$ lie in the Weyl alcove $\Sigma_0$ (i.e. the open simplex characterized by the conditions (i) and (ii) of Sect 2.3 with $g \downarrow 0$), then the only way in which this is possible is if $\sigma = id$ and $\mathbf{x} = \mathbf{y}$. (The Weyl alcove $\Sigma_0$ determines a fundamental domain for the action of the affine Weyl group $S_{N+1} \ltimes (\frac{2\pi}{\alpha} Q)$ on $E$.) The lemma then follows because the conditions on the parameters guarantee that $\rho + \Lambda_M^+ \subset \overline{\Sigma}_g \subset \Sigma_0$. □

After these preliminaries let us now introduce the wave function $\Psi_\lambda : (\rho + \Lambda_M^+) \to \mathbb{C}$ given by

$$(4.13) \qquad \Psi_\lambda(\rho + \mu) = \frac{1}{\mathcal{N}_0^{1/2}} \Delta^{1/2}(\lambda) \Delta^{1/2}(\mu) P_\lambda(\rho + \mu), \qquad \lambda, \mu \in \Lambda_M^+.$$

Notice that for $\lambda = \mathbf{0}$ this wave function reduces to the factorized wave function $\Psi_\mathbf{0}$ in (4.4). The next symmetry property is an immediate consequence of the symmetry relations (4.12) for the renormalized Macdonald polynomials $P_\lambda(\mathbf{x})$ (4.11).

**Proposition 4.6** (Symmetry). *One has that*

$$\Psi_\lambda(\rho+\mu) = \Psi_\mu(\rho+\lambda) \qquad for \quad \lambda,\mu \in \Lambda_M^+.$$

The function $\Psi_\lambda$ (4.13) turns out to be a joint eigenfunction of the operators $\hat{\mathcal{H}}_1, \ldots, \hat{\mathcal{H}}_N$.

**Proposition 4.7** (Diagonalization). *Let us assume positive parameters $\alpha, g$ subject to the constraint (3.5) and let $\lambda \in \Lambda_M^+$. Then*

$$\hat{\mathcal{H}}_r \Psi_\lambda = E_r(\rho+\lambda)\Psi_\lambda, \qquad r = 1, \ldots, N$$

*(where $\hat{\mathcal{H}}_r$ and $E_r(\cdot)$ are given by (3.1), (3.9) and (4.8), respectively, and $\rho$ is taken from (2.13)).*

*Proof.* One has that

$$(\hat{\mathcal{H}}_r^+ \Psi_\lambda)(\rho+\mu) \stackrel{(3.9)}{=} \sum_{\nu \in S_{N+1}(\omega_r)} W_\nu^+(\rho+\mu)\Psi_\lambda(\rho+\mu+\nu)$$

$$\stackrel{(i)}{=} \frac{1}{\mathcal{N}_0^{1/2}} \Delta^{1/2}(\lambda)\Delta^{1/2}(\mu) \times$$

$$\sum_{\nu \in S_{N+1}(\omega_r)} V_\nu(\rho+\mu) P_\lambda(\rho+\mu+\nu)$$

$$\stackrel{(ii)}{=} E_r^+(\rho+\lambda)\Psi_\lambda(\rho+\mu),$$

where we have used (i) Lemma 4.2 combined with the vanishing properties of the coefficients $W_\nu^+$ and $V_\nu$ at the boundary (cf. Lemma 3.2), and (ii) the defining difference equations for the Macdonald polynomials in (4.10b). The proposition now follows from the observation that $\hat{\mathcal{H}}_r = (\hat{\mathcal{H}}_r^+ + \hat{\mathcal{H}}_r^-)/2$ and that $E_r(\cdot) = (E_r^+(\cdot) + E_r^-(\cdot))/2$ with $\hat{\mathcal{H}}_r^- = \hat{\mathcal{H}}_{N+1-r}^+$ and $E_r^-(\cdot) = E_{N+1-r}^+(\cdot)$. □

It is clear from the proof of the above proposition that the functions $\Psi_\lambda$ (4.13) in fact diagonalize the operators $\hat{\mathcal{H}}_r^\pm$ (3.9) individually

$$(4.14) \qquad \hat{\mathcal{H}}_r^\pm \Psi_\lambda = E_r^\pm(\rho+\lambda)\Psi_\lambda, \qquad \lambda \in \Lambda_M^+$$

(with the parameters satisfying (3.5)).

It is a priori not obvious that the functions $\Psi_\lambda$, $\lambda \in \Lambda_M^+$ actually span the Hilbert space $L^2(\rho+\Lambda_M^+)$, since in principle linear dependencies might arise between the Macdonald polynomials $p_\lambda(\cdot)$, $\lambda \in \Lambda_M^+$ upon restriction to the lattice $\rho + \Lambda_M^+$. However, the following result states that the functions $\Psi_\lambda$, $\lambda \in \Lambda_M^+$ in fact form an *orthonormal basis* of $L^2(\rho + \Lambda_M^+)$, therewith excluding the possibility of such linear dependencies to occur. Phrased alternatively: the lattice evaluation homomorphism from the polynomial subspace $\text{Span}\{m_\lambda\}_{\lambda \in \Lambda_M^+}$ to the space of complex functions over the lattice $\rho + \Lambda_M^+$—defined by the assignment $m_\lambda(\mathbf{x}) \mapsto m_\lambda(\rho+\mu)$—is an isomorphism.

**Proposition 4.8** (Orthonormality). *For positive parameters $\alpha, g$ subject to condition (3.5), the functions $\Psi_\lambda : (\rho + \Lambda_M^+) \to \mathbb{C}$ in (4.13) form an orthonormal basis of $L^2(\rho + \Lambda_M^+)$, i.e.*

$$(\Psi_\lambda, \Psi_\mu) = \begin{cases} 0 & if \quad \lambda \neq \mu \\ 1 & if \quad \lambda = \mu \end{cases}$$

$(\lambda, \mu \in \Lambda_M^+)$.

*Proof.* The orthogonality follows by applying the eigenvalue equation (4.14) to the equality $(\hat{\mathcal{H}}_r^+ \Psi_\lambda, \Psi_\mu) = (\Psi_\lambda, \hat{\mathcal{H}}_r^- \Psi_\mu)$ (cf. the proof of Proposition 3.3) and using that the elementary symmetric functions $E_1^+(\cdot), \ldots, E_N^+(\cdot)$ separate the points of $\rho + \Lambda_M^+$ (cf. Lemma 4.5).

To see that the normalization of the wave functions is such that their $L^2$-norms are equal to 1, we apply the symmetry relations (Proposition 4.6) to the eigenvalue equations of Proposition 4.7. This leads to a system of difference equations for the wave functions in the spectral variable of the form

$$E_r(\rho + \lambda)\Psi_\mu(\rho + \lambda) = $$
$$\sum_{\nu \in S_{N+1}(\omega_r)} \Big( W_\nu^+(\rho + \mu)\Psi_{\mu+\nu}(\rho + \lambda) + W_\nu^-(\rho + \mu)\Psi_{\mu-\nu}(\rho + \lambda) \Big),$$

$r = 1, \ldots, N$ (where $W_\nu^\pm$ is taken from (3.9)). Applying the expansion on the r.h.s. to both sides of the equality $(E_r \Psi_\mu, \Psi_{\mu+\omega_r}) = (\Psi_\mu, E_r \Psi_{\mu+\omega_r})$ and exploiting the orthogonality of the wave functions, produces the relation

$$W_{\omega_r}^+(\rho + \mu)(\Psi_{\mu+\omega_r}, \Psi_{\mu+\omega_r}) = W_{\omega_r}^-(\rho + \mu + \omega_r)(\Psi_\mu, \Psi_\mu).$$

But then, since $W_{\omega_r}^+(\rho + \mu) = W_{\omega_r}^-(\rho + \mu + \omega_r) \; (> 0)$ for $\mu, \mu + \omega_r \in \Lambda_M^+$ (see (3.9)), it is immediate that $(\Psi_\mu, \Psi_\mu)$ is independent of $\mu \in \Lambda_M^+$. The orthonormality now follows because for $\mu = \mathbf{0}$ we have that $(\Psi_\mathbf{0}, \Psi_\mathbf{0}) = 1$ in view of Proposition 4.4. □

*Remarks:* i. The orthonormality of the wave functions $\Psi_\lambda$ (4.13) described by Proposition 4.8 can be rewritten in terms of discrete orthogonality relations for the Macdonald polynomials $p_\lambda(\mathbf{x})$ (in the monic normalization) or $P_\lambda(\mathbf{x})$ (in the symmetric normalization with $P_\lambda(\rho) = 1$, cf. (4.11), (4.12)). The discrete orthogonality measure is supported on the lattice $\rho + \Lambda_M^+$ with positive weights given by $\Delta$ (4.1). Specifically, we conclude from Proposition 4.7 that for positive parameters $\alpha, g$ subject to the condition (3.5) one has that

$$(4.15a) \qquad \sum_{\nu \in \Lambda_M^+} p_\lambda(\rho + \nu) \overline{p_\mu(\rho + \nu)} \Delta(\nu) = \begin{cases} 0 & if \quad \mu \neq \lambda \\ \mathcal{N}_0 \frac{C_-(\lambda)}{C_+(\lambda)} & if \quad \mu = \lambda, \end{cases}$$

or equivalently

$$(4.15b) \qquad \sum_{\nu \in \Lambda_M^+} P_\lambda(\rho + \nu) \overline{P_\mu(\rho + \nu)} \Delta(\nu) = \begin{cases} 0 & if \quad \mu \neq \lambda \\ \frac{\mathcal{N}_0}{\Delta(\lambda)} & if \quad \mu = \lambda, \end{cases}$$

for $\lambda, \mu \in \Lambda_M^+$ (with $C_\pm$ and $\mathcal{N}_0$ given by (4.2a), (4.2b) and Proposition 4.4, respectively).

*ii.* If we associate to each dominant weight vector $\lambda$ given by

$$\lambda = \sum_{j=1}^{N} l_j \,\omega_j, \qquad l_j \in \mathbb{N} \tag{4.16a}$$

a contragredient dominant weight $\lambda^*$ of the form

$$\lambda^* = \sum_{j=1}^{N} l_{N+1-j}\,\omega_j, \tag{4.16b}$$

then the mapping $\lambda \mapsto \lambda^*$ defines an involution of the cone of dominant weights $\Lambda^+$ (2.7). The Macdonald polynomials labeled by $\lambda$ and $\lambda^*$ are (for $\alpha, g$ real) related by complex conjugation

$$\overline{p_\lambda(\mathbf{x})} = p_{\lambda^*}(\mathbf{x}), \qquad \lambda \in \Lambda^+. \tag{4.17}$$

Indeed, the weight $-\lambda$ lies in the $S_{N+1}$-orbit of $\lambda^*$ and the vector $-\rho$ lies in the $S_{N+1}$-orbit of $\rho$, from which it is concluded that

$$\overline{m_\lambda(\mathbf{x})} = m_\lambda(-\mathbf{x}) = m_{\lambda^*}(\mathbf{x}), \qquad \overline{E_r^+(\rho+\lambda)} = E_r^-(\rho+\lambda) = E_r^+(\rho+\lambda^*).$$

Combining this with the observation that $\mu^* \prec \lambda^*$ if $\mu \prec \lambda$ then entails that the complex conjugate polynomial $\overline{p_\lambda(\mathbf{x})}$ satisfies the same conditions of the type in (4.10a), (4.10b) as the Macdonald polynomial $p_{\lambda^*}(\mathbf{x})$, whence the equality in (4.17) follows by the uniqueness of the Macdonald polynomials.

The upshot is that by passing to linear combinations of the form

$$\Psi_\lambda^C = \left(\frac{\Psi_\lambda + \overline{\Psi_\lambda}}{2}\right), \qquad \lambda \in \Lambda_M^+, \tag{4.18a}$$

$$\Psi_\lambda^S = \left(\frac{\Psi_\lambda - \overline{\Psi_\lambda}}{2i}\right), \qquad \lambda \in \Lambda_M^+ \tag{4.18b}$$

(thus selecting the real and imaginary parts of the wave functions $\Psi_\lambda$ (4.13)), we arrive at *real-valued* eigenfunctions for the discrete difference operators $\hat{\mathcal{H}}_1, \ldots, \hat{\mathcal{H}}_N$ (3.2a), (3.9)

$$\hat{\mathcal{H}}_r \Psi_\lambda^{C/S} = E_r(\rho+\lambda) \Psi_\lambda^{C/S}, \qquad \lambda \in \Lambda_M^+ \tag{4.19}$$

($r=1,\ldots,N$). Notice, however, that—in contrast to the complex wave functions $\Psi_\lambda$ (4.13)—the functions in (4.18a), (4.18b) do *not* diagonalize the operators $\hat{\mathcal{H}}_1^\pm, \ldots, \hat{\mathcal{H}}_N^\pm$ (3.9) individually. When $\lambda$ runs through the alcove $\Lambda_M^+$ (3.6), we of course count each eigenfunction $\Psi_\lambda^C$ and $\Psi_\lambda^S$ twice because

$$\Psi_{\lambda^*}^C = \Psi_\lambda^C, \qquad \Psi_{\lambda^*}^S = -\Psi_\lambda^S, \qquad \lambda \in \Lambda_M^+. \tag{4.20}$$

Eliminating for this redundancy yields a real-valued orthonormal basis for the Hilbert space $L^2(\rho+\Lambda_M^+)$, consisting of the wave functions $\Psi_\lambda^C$, $\Psi_\lambda^S$ with $\lambda \in \Lambda_M^+ \mod *$ (i.e., we pick the weights from a fundamental domain in $\Lambda_M^+$ with respect to the action of the involution $*$). The complex eigenbasis is recovered from the real eigenbasis by forming the combinations

$$\Psi_\lambda = \Psi_\lambda^C + i\Psi_\lambda^S, \qquad \lambda \in \Lambda_M^+. \tag{4.21}$$

## 5. Miscellanea

5.1. **The eigenfunction transform.** The weight alcove $\Lambda_M^+$ (3.6) labels the eigenbasis $\Psi_\lambda$ (4.13). If we identify this weight alcove with the lattice $\rho + \Lambda_M^+$ then we are led to a Discrete Fourier-type Transformation in the Hilbert space $L^2(\rho + \Lambda_M)$, the kernel of which is determined by the eigenfunctions.

Let $\mathcal{F} : L^2(\rho + \Lambda_M^+) \to L^2(\rho + \Lambda_M^+)$ be the discrete integral transformation

$$(5.1a) \qquad (\mathcal{F}f)(\rho + \lambda) = \sum_{\mu \in \Lambda_M^+} \mathcal{F}(\rho + \lambda, \rho + \mu) f(\rho + \mu)$$

with a kernel of the form

$$(5.1b) \qquad \mathcal{F}(\rho + \lambda, \rho + \mu) := \Psi_\lambda(\rho + \mu),$$

where $\Psi_\lambda(\rho + \mu)$ is taken from (4.13). Furthermore, let $\mathcal{E}_r : L^2(\rho + \Lambda_M^+) \to L^2(\rho + \Lambda_M^+)$ denote the multiplication operator

$$(5.2) \qquad (\mathcal{E}_r f)(\rho + \mu) = E_r(\rho + \mu)\, f(\rho + \mu) \qquad (r = 1, \dots, N)$$

with $E_r(\cdot)$ representing the real elementary symmetric function of (4.8).

The main results of this paper may be conveniently summarized in the following three properties of the discrete integral transformation $\mathcal{F}$ (5.1a), (5.1b)

$$(5.3a) \qquad {}^t\mathcal{F} = \mathcal{F}, \qquad \mathcal{F}^* = \mathcal{F}^{-1}$$

and

$$(5.3b) \qquad \mathcal{F}\hat{\mathcal{H}}_r = \mathcal{E}_r \mathcal{F}, \qquad r = 1, \dots, N,$$

where it is assumed that the parameters satisfy condition (3.5). The first property states that the transpose ${}^t\mathcal{F}$ of $\mathcal{F}$ is equal to $\mathcal{F}$, or in other words, that (the kernel of) the operator $\mathcal{F}$ is symmetric. This is a consequence of the symmetry relation in Proposition 4.6. The second property states that the adjoint $\mathcal{F}^*$ of $\mathcal{F}$ in $L^2(\rho + \Lambda_M^+)$ equals the inverse of $\mathcal{F}$, or in other words, that the operator $\mathcal{F}$ is unitary. This is a consequence of the orthonormality relations for the kernel $\Psi_\lambda(\rho + \mu)$ in Proposition 4.8. Finally, the third property states that $\mathcal{F}$ simultaneously diagonalizes the discrete difference operators $\hat{\mathcal{H}}_1, \dots, \hat{\mathcal{H}}_N$ (3.1), (3.9) in $L^2(\rho + \Lambda_M^+)$. This is seen by checking that both sides of (5.3b) act the same on the orthonormal eigenbasis $\overline{\Psi}_\lambda$, $\lambda \in \Lambda_M^+$ (cf. Proposition 4.7 and also Remark *ii.* at the end of Sect. 4).

The map $f \mapsto \hat{f} := \mathcal{F}^* f$ determines a Discrete Fourier-type Transformation in $L^2(\rho + \Lambda_M^+)$ of the form

$$(5.4a) \qquad \hat{f}(\rho + \lambda) = (f, \Psi_\lambda), \qquad \lambda \in \Lambda_M^+$$

with the inversion formula given by

$$(5.4b) \qquad f(\rho + \mu) = (\hat{f}, \overline{\Psi}_\mu), \qquad \mu \in \Lambda_M^+.$$

The coefficients $\hat{f}_\lambda := \hat{f}(\rho + \lambda)$, $\lambda \in \Lambda_M^+$ solve the linear interpolation problem of decomposing an arbitrary lattice function $f \in L^2(\rho + \Lambda_M^+)$ in terms of the wave functions $\Psi_\lambda$, $\lambda \in \Lambda_M^+$

$$\tag{5.5} f = \sum_{\lambda \in \Lambda_M^+} \hat{f}_\lambda \Psi_\lambda, \qquad \hat{f}_\lambda = (f, \Psi_\lambda).$$

By passing to the real eigenbasis from Remark *ii.* at the end of Sect 4, which is given by $\Psi_\lambda^C$, $\Psi_\lambda^S$ (4.18a), (4.18b) with $\lambda \in \Lambda_M^+ \mod *$, we arrive at analogs of the (discrete) Fourier cosine transform $\mathcal{F}^c$

$$\tag{5.6a} \hat{f}^c(\rho + \lambda) = (f, \Psi_\lambda^C), \qquad \lambda \in \Lambda_M^+ \mod *$$

$$\tag{5.6b} f(\rho + \mu) = (\hat{f}^c, \Psi_\mu^C), \qquad \mu \in \Lambda_M^+ \mod *$$

and Fourier sine transform $\mathcal{F}^s$

$$\tag{5.7a} \hat{f}^s(\rho + \lambda) = (f, \Psi_\lambda^S), \qquad \lambda \in \Lambda_M^+ \mod *$$

$$\tag{5.7b} f(\rho + \mu) = (\hat{f}^s, \Psi_\mu^S), \qquad \mu \in \Lambda_M^+ \mod *,$$

respectively. Here it is assumed that $f \in L^2(\rho + \Lambda_M^+)$ is a lattice function that is symmetric (for the cosine transform) or antisymmetric (for the sine transform) with respect to the action of the involution $*$ on the lattice $\rho + \Lambda_M^+$, respectively. (The involution $*$ acts on the lattice $\rho + \Lambda_M^+$ by $(\rho + \lambda) \mapsto (\rho + \lambda^*)$, cf. Remark *ii.* at the end of Sect. 4.)

The integral transformation $\mathcal{F}$ (5.1a), (5.1b) conjugates (cf. (5.3b)) the quantum Hamiltonians $\hat{\mathcal{H}}_1, \ldots, \hat{\mathcal{H}}_N$ (3.1), (3.9) to the multiplication operators $\mathcal{E}_1, \ldots, \mathcal{E}_N$ (5.2). The spectrum of $\hat{\mathcal{H}}_r$ in $L^2(\rho + \Lambda_M^+)$ is (thus) given by the range of the real elementary symmetric function $E_r(\cdot) = \sum_{\nu \in S_{N+1}(\omega_r)} \cos \alpha \langle \nu, \cdot \rangle$ on the lattice $\rho + \Lambda_M^+$. The eigenfunction transform $\mathcal{F}$ amounts to the quantum counterpart of the action-angle transformation $\phi$ for the classical system, which was found in explicit form by Ruijsenaars [R3].

The action-angle transform at issue constitutes an (anti)symplectomorphism $\mathbf{z} \xrightarrow{\phi} \check{\mathbf{z}}$ of the classical phase space $(\mathbb{CP}^N, \omega_R)$, giving rise to new canonical coordinates $(\check{\mathbf{x}}, \check{\mathbf{p}})$ of the form (cf. (2.19a), (2.19b))

$$\tag{5.8a} \langle \mathbf{a}_j, \check{\mathbf{p}} \rangle = (2\pi/\alpha - (N+1)g) \frac{|\check{z}_j|^2}{|\check{z}_0|^2 + \cdots + |\check{z}_N|^2} + g,$$

$$\tag{5.8b} e^{i \langle \omega_j, \check{\mathbf{x}} \rangle} = \frac{\check{z}_j |\check{z}_0|}{\check{z}_0 |\check{z}_j|}$$

($j = 1, \ldots, N$) on an open dense patch $\{[\check{z}_0 : \cdots : \check{z}_N] \mid \check{z}_j \neq 0, \ j = 0, \ldots, N\}$ of $\mathbb{CP}^N$. The classical Hamiltonians $\mathcal{H}_r(\mathbf{x}, \mathbf{p})$ (2.15) become in these new coordinates of the form [R3]

$$\tag{5.9} \check{\mathcal{H}}_r(\check{\mathbf{x}}, \check{\mathbf{p}}) = \sum_{\nu \in S_{N+1}(\omega_r)} \cos \alpha \langle \nu, \check{\mathbf{p}} \rangle, \qquad r = 1, \ldots, N.$$

(This is to be compared with the $\mathcal{F}$-transformation of the quantum Hamiltonian $\hat{\mathcal{H}}_r$ to the multiplication operator $\mathcal{E}_r$ in (5.3b).) That is, in the new coordinates the classical Hamiltonians depend only on the action variables $\check{\mathbf{p}} \in \Sigma_g$ and are independent of the angle variables $\check{\mathbf{x}} \in T = E/(2\pi Q)$ (cf. Sect. 2.3). Thus, the "spectrum" (i.e. the range) of the classical Hamiltonian $\mathcal{H}_r$ (2.15) on the compactified phase space $\mathbb{CP}^N$ is given by the range of $\check{\mathcal{H}}_r$ (5.9) on the convex polytope $\{\check{\mathbf{p}} \mid \check{\mathbf{p}} \in \overline{\Sigma}_g\}$ (cf. (2.22)). (Notice that we may again continue the action variables $\check{\mathbf{p}}$ (5.8a) smoothly to the whole of $\mathbb{CP}^N$ unlike the angle variables $\check{\mathbf{x}}$ (5.8b), cf. Sect. 2.3.) At this point it is worthwhile to mention that the convexity of the range of the action variables for our model is in agreement with the general convexity results for Hamiltonian systems on compact symplectic manifolds due to Atiyah [At] and Guillemin-Sternberg [GS1]. We see that—roughly speaking—the quantization of the model discretizes the spectrum of the Hamiltonians as if the action variables $\check{\mathbf{p}}$ get localized on the lattice $\rho + \Lambda_M^+ \subset \overline{\Sigma}_g$. It was furthermore shown by Ruijsenaars [R3], that the vertices $\check{\mathbf{p}} = \rho$, $\rho + M\omega_r$ ($r = 1, \ldots, N$) of the convex polytope $\{\check{\mathbf{p}} \mid \check{\mathbf{p}} \in \overline{\Sigma}_g\}$ correspond to the equilibrium points of the flows generated by the classical Hamiltonians $\mathcal{H}_r$ (2.15) (cf. also Sect. 5.2 below). This state of affairs is again in agreement with the general picture presented by Atiyah and Guillemin-Sternberg [At, GS1].

A remarkable property of the action-angle transform $\phi$ is that it in fact defines an (anti)symplectic *involution* on $(\mathbb{CP}^N, \omega_R)$ (i.e. the model is "self-dual" in the terminology of Ruijsenaars) [R3]. The analogous property of the eigenfunction transform $\mathcal{F}$ (5.1a), (5.1b) states that the corresponding discrete integral transform defines a (discrete) Fourier-type involution (i.e. an involution up to complex conjugation) in the Hilbert space $L^2(\rho + \Lambda_M^+)$

$$\text{(5.10)} \qquad {}^t\mathcal{F}^* \mathcal{F} = Id,$$

cf. (5.3a). (The kernel of $\mathcal{F}$ and ${}^t\mathcal{F}^*$ are the same up to complex conjugation.)

### 5.2. Ground-state vs maximal energy wave function. Let

$$\text{(5.11)} \qquad E(\mathbf{x}) = \sum_{j-1}^{N+1} \cos(\alpha x_j).$$

Notice that on the hyperplane $x_1 + \cdots + x_{N+1} = 0$ this function coincides with the first (real) elementary symmetric function $E_1(\mathbf{x}) = \sum_{\nu \in S_{N+1}(\omega_1)} \cos \alpha \langle \nu, \mathbf{x} \rangle$ (cf. (4.8)). It is not very difficult to infer that the critical points of $E(\mathbf{x})$ on the simplex $\overline{\Sigma}_g$ (2.22) are located at the $N+1$ vertices

$$\text{(5.12)} \qquad \rho \quad \text{and} \quad \rho + M\omega_r, \quad r = 1, \ldots, N,$$

where $M = \frac{2\pi}{\alpha} - (N+1)g$. The (critical) values of $E(\mathbf{x})$ evaluated at these vertices read (cf. (4.6))

$$\text{(5.13a)} \qquad E(\rho) = \frac{\sin \frac{\alpha g}{2}(N+1)}{\sin(\frac{\alpha g}{2})},$$

$$\text{(5.13b)} \qquad E(\rho + M\omega_r) = \cos(\frac{2\pi r}{N+1}) E(\rho), \qquad r = 1, \ldots, N.$$

We thus conclude (cf. Proposition 4.7) that the maximal eigenvalue of the Hamiltonian $\hat{\mathcal{H}}_1$ (3.1), (3.9) in the Hilbert space $L^2(\rho + \Lambda_M^+)$ has the value $E_1(\rho) =$ $\frac{\sin\frac{\alpha}{2}(N+1)g}{\sin(\frac{\alpha g}{2})}$. The corresponding eigenfunction is given by the factorized wave function $\Psi_\mathbf{0}$ in (4.4). For $N$ odd the minimal eigenvalue reads $E_1(\rho + M\omega_{(N+1)/2}) = -E_1(\rho)$ whereas for $N$ even the minimal eigenvalue is twofold degenerate and given by $E_1(\rho + M\omega_{N/2}) = E_1(\rho + M\omega_{(N/2+1)}) = \cos(\frac{\pi N}{N+1})E_1(\rho)$.

The "critical" or "vertex" eigenvalues of the Hamiltonian $\hat{\mathcal{H}}_1$ in (5.13a) and (5.13b) coincide with the equilibrium values of the corresponding classical Hamiltonian $\mathcal{H}_1$ (2.15) at the stationary points computed by Ruijsenaars [R3, Sect. 5.3] (cf. Sect. 5.1 above). In particular, the global minimal and maximal energies of the Hamiltonian $\hat{\mathcal{H}}_1/\mathcal{H}_1$ read the same at the quantum level as they do at the classical level. In other words, the energy levels get discretized at the quantum level (cf. Sect. 5.1 above) but there is no shift of the energy spectrum due to the quantization (as e.g. in the case of a harmonic oscillator).

From a physical point of view it is often somewhat more natural to work with a nonnegative Hamiltonian. This can be achieved by passing to difference operators of the form

$$\begin{aligned}
(5.14) \quad \tilde{\mathcal{H}}_r &= E_r(\rho) - \hat{\mathcal{H}}_r \\
&= \frac{1}{2} \sum_{\nu \in S_{N+1}(\omega_r)} \Big( V_\nu(\mathbf{x}) + V_\nu(-\mathbf{x}) \\
&\quad - V_\nu^{1/2}(\mathbf{x}) T_\nu V_\nu^{1/2}(-\mathbf{x}) - V_\nu^{1/2}(-\mathbf{x}) T_\nu^{-1} V_\nu^{1/2}(\mathbf{x}) \Big),
\end{aligned}$$

$r = 1, \ldots, N$ (where we have again employed the Macdonald identity from the proof of Proposition 4.3 to pass from the first to the second formula on the r.h.s.). The factorized eigenfunction $\Psi_\mathbf{0}$ (4.4) amounts to the ground-state wave function for the Hamiltonian $\tilde{\mathcal{H}}_1$ (5.14), with the corresponding eigenvalue being equal to zero.

5.3. **The two-particle solution.** In the case of two particles, i.e. for $N = 1$, the quantum version of the compactified Ruijsenaars-Schneider model was introduced and solved already several years ago by Ruijsenaars in the survey paper [R2] (see Sect. 3C2). It is instructive to view how, in this special situation, our results reproduce those previously obtained by Ruijsenaars. The difference operator $\hat{\mathcal{H}}_1$ (3.1) ($= \hat{\mathcal{H}}_1^+$ (3.2a) $= \hat{\mathcal{H}}_1^-$ (3.2b)) reduces to

$$(5.15) \quad \hat{\mathcal{H}} = \left(\frac{\sin\frac{\alpha}{2}(x+g)}{\sin(\frac{\alpha x}{2})}\right)^{1/2} e^{\frac{d}{dx}} \left(\frac{\sin\frac{\alpha}{2}(x-g)}{\sin(\frac{\alpha x}{2})}\right)^{1/2} +$$
$$\left(\frac{\sin\frac{\alpha}{2}(x-g)}{\sin(\frac{\alpha x}{2})}\right)^{1/2} e^{-\frac{d}{dx}} \left(\frac{\sin\frac{\alpha}{2}(x+g)}{\sin(\frac{\alpha x}{2})}\right)^{1/2}$$

with $x = x_1 - x_2$. For

$$(5.16) \quad \alpha, g > 0 \quad \text{and} \quad \frac{2\pi}{\alpha} - 2g = M \in \mathbb{N} \setminus \{0\},$$

the operator $\hat{\mathcal{H}}$ (5.15) becomes self-adjoint upon restriction to the Hilbert space $L^2(g + \Lambda_M^+)$ over the lattice

$$(5.17) \qquad g + \Lambda_M^+ = \{g, g+1, g+2, \ldots, g+M\}$$

(cf. Sect. 3.2). An orthonormal basis of eigenfunctions $\Psi_l : (g + \Lambda_M^+) \to \mathbb{R}$ for $\hat{\mathcal{H}}$ (5.15) is given by (cf. Sect. 4, in particular Eqs. (4.13), (4.1) and Proposition 4.4)

$$(5.18) \qquad \Psi_l(g+m) = \frac{1}{\mathcal{N}_0^{1/2}} \Delta^{1/2}(l) \Delta^{1/2}(m) \, P_l(\cos\frac{\alpha}{2}(g+m)),$$

$l, m = 0, \ldots, M$, with

$$\Delta(m) = \frac{\sin\frac{\alpha}{2}(m+g)}{\sin(\frac{\alpha g}{2})} \frac{(2g : \sin_\alpha)_m}{(1 : \sin_\alpha)_m}$$

$$= \frac{\sin\frac{\alpha}{2}(m+g)}{\sin(\frac{\alpha g}{2})} \prod_{j=1}^m \frac{\sin\frac{\alpha}{2}(2g+j-1)}{\sin(\frac{\alpha j}{2})},$$

$$P_l(\cos\frac{\alpha}{2}(x)) = q^{l(x-g)/2} {}_3\phi_2\left(\begin{matrix} q^{-l},\, q^g,\, q^{g-x} \\ q^{2g},\, 0 \end{matrix} ; q, q\right), \qquad q = e^{i\alpha}$$

and

$$\mathcal{N}_0 = 2^M (1+g : \sin_\alpha)_{M-1} = 2^M \prod_{k=1}^{M-1} \sin\frac{\alpha}{2}(k+g).$$

Specifically, one has that (cf. Proposition 4.7)

$$(5.19) \qquad \hat{\mathcal{H}} \Psi_l = 2\cos\frac{\alpha}{2}(g+l) \, \Psi_l, \quad l = 0, \ldots, M$$

and that (cf. Proposition 4.8)

$$(5.20) \qquad \sum_{m=0}^M \Psi_l(g+m) \overline{\Psi_k(g+m)} = \delta_{l,k}, \quad l, k \in \{0, \ldots, M\}.$$

Clearly, the maximal and minimal eigenvalues of the Hamiltonian $\hat{\mathcal{H}}$ are given by $2\cos(\frac{\alpha g}{2})$ and $2\cos\frac{\alpha}{2}(g+M) = -2\cos(\frac{\alpha g}{2})$, respectively (cf. Sect. 5.2).

The ($A_1$-type Macdonald) polynomials $P_l(\cos\frac{\alpha}{2}(x))$ coincide up to a normalization factor with the $q$-ultraspherical polynomials and can thus be explicitely written (in various ways) in terms of terminating basic hypergeometric series [GR, KS]. For our purposes it is convenient to employ the above representation in terms of a terminating ${}_3\phi_2$ series as this manifestly demonstrates the symmetry of the wave function $\Psi_l(g+m)$ (5.18) with respect to an interchange of $l$ and $m$ (cf. Proposition 4.6). The orthogonality relations (5.20) for the basis $\Psi_l$, $l = 0, \ldots, M$ boil down to reductions to special parameter values of well-known discrete orthogonality relations for the $q$-Racah polynomials due to Askey and Wilson [AS, GR, KS].

Notice that the rank-one case $N = 1$ is very special in the sense that the reflection $\lambda \mapsto -\lambda$ lies in the Weyl group (it amounts to the Weyl-permutation $(x_1, x_2) \mapsto (x_2, x_1)$ restricted to the hyperplane $E = \{(x_1, x_2) \in \mathbb{R}^2 \mid x_1 + x_2 = 0\}$). As a consequence, the involution $*$ of Remark $ii$ at the end of Sect. 4 reduces in this special case to the identity. Indeed, we have for $N = 1$ that the wave function $\Psi_\lambda$ (4.13) is real-valued, whence $\Psi_\lambda = \Psi_\lambda^C$ (4.18a) and $\Psi_\lambda^S = 0$ (4.18b). In other words, the "Discrete Fourier transform" $\mathcal{F} = \mathcal{F}^*$ (5.4a), (5.4b) and the "Discrete Fourier cosine transform" $\mathcal{F}^c$ (5.6a), (5.6b) coincide in the rank-one case and the "Discrete Fourier sine transform" $\mathcal{F}^s$ (5.7a), (5.7b) collapses.

5.4. **Geometric quantization.** In the light of the fact that the phase space for the classical compactified Ruijsenaars-Schneider model is given by the complex projective space $\mathbb{CP}^N$ equipped with the renormalized Fubini-Study symplectic form $\omega_R$ (2.20), it is natural to ask oneself the question as to what extent our results may be recovered within the realms of geometric quantization. In this formalism a Hilbert space is associated to the classical phase space $(\mathbb{CP}^N, \omega_R)$ in two steps (see e.g. [Si, Hu]). In the first step (prequantization), the question is to construct a Hermitian line bundle $\mathbb{L}$ with connection $\nabla$ over $\mathbb{CP}^N$ such that the curvature of $\nabla$ equals $\omega_R$. Such a line bundle exists provided $\omega_R$ satisfies the integrality condition

$$(5.21) \qquad \oint \omega_R = 2\pi M \quad \text{with } M \in \mathbb{N} \setminus \{0\},$$

where the integration is over a complex projective line in $\mathbb{CP}^N$ (or more generally an integral two-cycle in the homology basis). (Geometrically, this condition means that $\omega_R$ belongs to an integer cohomology class: $[\omega_R] \in H^2(\mathbb{CP}^N, \mathbb{Z})$.) The prequantum Hilbert space now consists of the space of $L^2$ sections of the line bundle $\mathbb{L}$ (where the measure of integration is taken to be the Liouville volume form associated to $\omega_R$). After recalling that the normalization of $\omega_R$ is such that $\oint \omega_R = 4\pi R^2$ (2.21), it is seen that the integrality condition in (5.21) amounts precisely to our quantization condition (3.5). (This observation was already made by Ruijsenaars in [R3, Sect. 1.3].)

Unfortunately, the (prequantum) Hilbert space thus obtained is too big. Roughly speaking, it corresponds to an "$L^2$ space over the phase space" whereas from a physical point of view one is interested rather in the analog of an "$L^2$ space over the configuration space". In the second step of the quantization procedure the prequantum Hilbert space has to be downsized so as to produce the physical Hilbert space. To this end it is needed to exploit the fact that $(\mathbb{CP}^N, \omega_R)$ is a Kähler manifold and, as such, carries a natural Kähler polarization. Specifically, as the physical Hilbert space one picks the subspace of the prequantum Hilbert space consisting of all $L^2$ sections of the line bundle $\mathbb{L}$ that are covariantly constant (with respect to the connection $\nabla$) along the leaves of the (standard) Kähler polarization on $\mathbb{CP}^N$. The result is a physical Hilbert space $\mathcal{H}_{hol}$ consisting of the holomorphic sections of $\mathbb{L}$. The dimension of the space of holomorphic sections $\mathcal{H}_{hol}$ follows from a classical result (viz. a Riemann-Roch-type formula) due to Hirzebruch and Kodaira [HK]

$$(5.22) \qquad \dim(\mathcal{H}_{hol}) = \frac{(N+M)!}{N!\,M!},$$

which corresponds nicely to the dimension of our Hilbert space $L^2(\rho + \Lambda_M^+)$ in (3.8). It is possible to realize the Hilbert space $\mathcal{H}_{hol}$ more explicitly, as the space of holomorphic sections may be identified with the space of functions of the form [HK, Si, Hu]

$$(5.23) \qquad \frac{p(z_1, \ldots, z_N)}{(1 + |z_1|^2 + \cdots + |z_N|^2)^M},$$

where $p(z_1, \ldots, z_N)$ denotes an arbitrary polynomial of degree at most $M$ in the affine $\mathbb{CP}^N$ coordinates $(z_1, \cdots, z_N)$. In this representation the integration of the $L^2$ inner product is with respect to the volume form

$$(5.24) \qquad (1 + |z_1|^2 + \cdots + |z_N|^2)^{-(N+1)} dz_1 \wedge d\overline{z}_1 \wedge \cdots \wedge dz_N \wedge d\overline{z}_N.$$

It would be very interesting to extend the analysis further so as to include a description of the quantum Hamiltonians and their eigenfunctions within the framework of geometric quantization and to compare the results with the approach taken in the present paper. In this connection it is expected that our lattice $\rho + \Lambda_M^+$ may be recovered geometrically as the so-called Bohr-Sommerfeld set—see Guillemin-Sternberg [GS2]—associated to the symplectic embedding of $\Sigma_g \times T$ into $\mathbb{CP}^N$ given by (2.18). Furthermore, the embedding in question induces a "real polarization with singularities" on $\mathbb{CP}^N$ (cf. [GS2]). (This real polarization becomes singular at the boundary hyperplanes $z_j = 0$ of the open dense patch $\{[z_0 : \cdots : z_N] \mid z_j \neq 0,\ j = 0, \ldots, N\} \subset \mathbb{CP}^N$.) The above-mentioned correspondence between (the dimensions of) the Hilbert space $L^2(\rho + \Lambda_M^+)$ and the Hilbert space of holomorphic sections of the line bundle $\mathbb{L}$ (viz. $\mathcal{H}_{hol}$) suggests that (the dimension of) the Hilbert space associated to $(\mathbb{CP}^N, \omega_R)$ via geometric quantization does not depend on the choice of polarization to be either the standard Kähler polarization or the "real polarization with singularities" stemming from the embedding $\Sigma_g \times T \hookrightarrow \mathbb{CP}^N$. This is in correspondence with the more general "invariance of polarization" results for the geometric quantization of complex flag manifolds due to Guillemin and Sternberg [GS2].

5.5. **Connections to integrable field theories.** The compactified Ruijsenaars-Schneider model is related in various ways to well-known infinite-dimensional integrable systems. For instance, in [R5] it was shown that at the classical level the ($\tau$-functions of) single-solitons for the Kadomtsev-Petviashvili and $2D$ Toda hierarchy (cf. [DKJM, JM, Ho1, Ho2, ZC]) may be described in terms of the equilibrium behavior of the classical compactified Ruijsenaars-Schneider molecule with an appropriate center-of-mass motion. Multi-solitons arise in this picture—in a nutshell—by passing to composite integrable Ruijsenaars-Schneider-type $(m_1 + \cdots + m_n)$-particle systems that are built of $n$ (the number of solitons) interacting compactified trigonometric Ruijsenaars-Schneider molecules in their ground state [R5].

In [GN] it was furthermore argued that formally the (quantum) compactified Ruijsenaars-Schneider model can be obtained by Hamiltonian reduction from an infinite-dimensional system on the cotangent bundle over a central extension of the loop group $\widehat{SU}(N+1)$. The latter paper also indicates some intriguing relations between, on the one hand, the quantum compactified Ruijsenaars-Schneider model and, on the other hand, a gauged $SU(N+1)/SU(N+1)$ Wess-Zumino-Witten topological quantum field theory on a cylinder and a Chern-Simons theory with gauge group $SU(N+1)$ on a three-fold that is the product of an interval and a real two-torus.

5.6. **Bispectrality.** The symmetry of the wave function $\Psi_\lambda(\rho+\mu)$ (4.13) with respect to an interchange of $\lambda$ and $\mu$ (cf. Proposition 4.6) has as consequence that it satisfies the same discrete difference eigenvalue equations in the "spectral" variable $\lambda$ as it does in the "spatial" variable $\mu$ (cf. Proposition 4.7 and the proof of Proposition 4.8). In other words, we are dealing with a multivariate doubly discrete finite-dimensional bispectral problem in the sense of Duistermaat and Grünbaum [DG, W, G].

APPENDIX A. TRUNCATED AND TERMINATING AOMOTO-ITO-MACDONALD SUMS

In this appendix a finitely truncated version of a recent summation formula due to Aomoto, Ito, and Macdonald [Ao, I, M5] (see also [Ka]) is derived. In Sect. 4 we used this finite Aomoto-Ito-Macdonald-type sum to arrive at a compact product formula for the normalization constant of our factorized wave function $\Psi_0$ (4.4) (cf. Proposition 4.4). When formulating the summation formulas in question it is convenient to employ the $q$-shifted factorial defined by (see e.g. [GR])

$$(a;q)_m = \begin{cases} (1-aq)(1-aq^2)\cdots(1-aq^{m-1}), & m=1,2,3,\ldots \\ 1, & m=0 \\ \frac{1}{(1-aq^{-1})(1-aq^{-2})\cdots(1-aq^m)}, & m=-1,-2,-3,\ldots \end{cases}$$

(where for negative $m$ it is assumed that $a,q \in \mathbb{C}$ are such that the denominators do not vanish),

$$(a;q)_\infty = \prod_{n=1}^\infty (1-aq^n), \qquad (0<|q|<1)$$

and

$$(a_1,\ldots,a_k;q)_m = (a_1;q)_m \cdots (a_k;q)_m, \qquad m \in \mathbb{Z} \cup \{\infty\}.$$

**Proposition A.1** (The Aomoto-Ito-Macdonald sum [Ao, I, M5])**.** *For $0<q<1$, $Re(g)<0$ and $\mathbf{z} \in \mathbb{C}^{N+1}$ with $g+\langle \mathbf{a},\mathbf{z}\rangle \notin \mathbb{Z} + \frac{2\pi}{i\log q}\mathbb{Z}$ for all $\mathbf{a} \in A_N^+$, one has that*

$$\sum_{\mu \in \Lambda} q^{-2\langle \rho, \mathbf{z}+\mu\rangle} \prod_{\mathbf{a}\in A_N^+} (1-q^{\langle \mathbf{a},\mathbf{z}+\mu\rangle}) \frac{(q^{1-g+\langle \mathbf{a},\mathbf{z}+\mu\rangle};q)_\infty}{(q^{g+\langle \mathbf{a},\mathbf{z}+\mu\rangle};q)_\infty} = \gamma\,\Theta(\mathbf{z})$$

*(where the sum is taken over all weights in $\Lambda$ (2.5) and the vector $\rho$ is given by (2.13))*, with

$$\Theta(\mathbf{z}) = q^{-2\langle \rho,\mathbf{z}\rangle} \prod_{\mathbf{a}\in A_N^+} \frac{\theta(q^{\langle \mathbf{a},\mathbf{z}\rangle})}{\theta(q^{g+\langle \mathbf{a},\mathbf{z}\rangle})}, \qquad \theta(\zeta) = (q,\zeta,q\zeta^{-1};q)_\infty$$

and

$$\gamma = (N+1) \prod_{\mathbf{a}\in A_N^+} \frac{(q^{1-g-\langle \mathbf{a},\rho\rangle},q^{\delta_\mathbf{a}+g-\langle \mathbf{a},\rho\rangle};q)_\infty}{(q^{1-\langle \mathbf{a},\rho\rangle},q^{-\langle \mathbf{a},\rho\rangle};q)_\infty},$$

where $\delta_{\mathbf{a}} := 1$ if $\mathbf{a}$ is simple (cf. (2.2)) and $\delta_{\mathbf{a}} := 0$ otherwise. Furthermore, the series on the l.h.s. converges in absolute value.

The conditions on $q$, $g$ ensure that the series on the l.h.s. converges absolutely and the genericity restrictions on $\mathbf{z}$ guarantee that all denominators are nonzero. The sum of Proposition A.1 was first considered by Aomoto [Ao], who showed that it can be evaluated as a product of the quasi-periodic factor $\Theta(\mathbf{z})$ and a $\mathbf{z}$-independent constant. The value of this constant (viz. $\gamma$), was subsequently conjectured by Ito [I]. In its present form the statement of the above proposition is due to Macdonald [M5], who derived it by linking a constant term identity of Cherednik [C1] to a generalized Poincaré series type formula for affine Weyl groups due to Matsumoto [Ma].

After division of both sides of the Aomoto-Ito-Macdonald summation formula by the middle term on the l.h.s. corresponding to $\mu = \mathbf{0}$, one arrives at the identity

$$\text{(A.1)} \quad \sum_{\mu \in \Lambda} q^{-2\langle \rho, \mu \rangle} \prod_{\mathbf{a} \in A_N^+} \Big( \frac{1 - q^{\langle \mathbf{a}, \mathbf{z} + \mu \rangle}}{1 - q^{\langle \mathbf{a}, \mathbf{z} \rangle}} \Big) \frac{(q^{g + \langle \mathbf{a}, \mathbf{z} \rangle}; q)_{\langle \mathbf{a}, \mu \rangle}}{(q^{1-g + \langle \mathbf{a}, \mathbf{z} \rangle}; q)_{\langle \mathbf{a}, \mu \rangle}}$$
$$= \gamma \prod_{\mathbf{a} \in A_N^+} \frac{(q^{1+\langle \mathbf{a}, \mathbf{z} \rangle}, q^{1-\langle \mathbf{a}, \mathbf{z} \rangle}; q)_\infty}{(q^{1-g+\langle \mathbf{a}, \mathbf{z} \rangle}, q^{1-g-\langle \mathbf{a}, \mathbf{z} \rangle}; q)_\infty},$$

where it is assumed that $0 < q < 1$, $Re(g) < 0$ and that $\mathbf{z} \in \mathbb{C}^{N+1}$ satisfies the genericity conditions $\langle \mathbf{a}, \mathbf{z} \rangle \notin \frac{2\pi}{i \log(q)} \mathbb{Z}$ and $g + \langle \mathbf{a}, \mathbf{z} \rangle - 1 \notin \mathbb{N} + \frac{2\pi}{i \log(q)} \mathbb{Z}$ for all $\mathbf{a} \in A_N$ (to ensure that there is no division by zero). It is instructive to observe that the proportionality constant $\gamma$ (see above) on the r.h.s. may be rewritten in a somewhat more compact (but less elegant) form by canceling common factors in the numerator and denominator:

$$\text{(A.2)} \quad \gamma = (N+1) \frac{(q; q)_\infty^N}{(q^{1-g}; q)_\infty^N} \prod_{1 \leq n \leq N} \frac{(q^{1-(N+1)g}; q)_\infty}{(q^{-ng}; q)_\infty}.$$

We will now show that by specializing the vector $\mathbf{z}$ to the value $\rho$ (2.13), the sum in (A.1) over the weight lattice $\Lambda$ (2.5) truncates to a sum over the dominant cone $\Lambda^+$ (2.7).

**Proposition A.2** (A Truncated Aomoto-Ito-Macdonald sum). *Let $0 < q < 1$ and $Re(g) < 0$ such that $g - \langle \mathbf{a}, \rho \rangle$ is not a positive integer modulo $\frac{2\pi}{i \log(q)}$ for all $\mathbf{a} \in A_N^+$. Then*

$$\sum_{\mu \in \Lambda^+} q^{-2\langle \rho, \mu \rangle} \prod_{\mathbf{a} \in A_N^+} \Big( \frac{1 - q^{\langle \mathbf{a}, \rho + \mu \rangle}}{1 - q^{\langle \mathbf{a}, \rho \rangle}} \Big) \frac{(q^{g+\langle \mathbf{a}, \rho \rangle}; q)_{\langle \mathbf{a}, \mu \rangle}}{(q^{1-g+\langle \mathbf{a}, \rho \rangle}; q)_{\langle \mathbf{a}, \mu \rangle}}$$
$$= (N+1) \prod_{\mathbf{a} \in A_N^+} \frac{(q^{1+\langle \mathbf{a}, \rho \rangle}, q^{\delta_{\mathbf{a}} + g - \langle \mathbf{a}, \rho \rangle}; q)_\infty}{(q^{-\langle \mathbf{a}, \rho \rangle}, q^{1-g+\langle \mathbf{a}, \rho \rangle}; q)_\infty}$$
$$= (N+1) \prod_{1 \leq n \leq N} \frac{(q^{1+ng}; q)_\infty}{(q^{-ng}; q)_\infty}$$

*and the series on the l.h.s. converges in absolute value.*

*Proof.* The conditions on $g$ (and $q$) ensure that after substituting $\mathbf{z} = \rho$ (2.13) in (A.1) all terms remain finite and the series converges in absolute value (as a consequence of Proposition A.1). The resulting series on the l.h.s. now truncates because the terms become zero for $\mu \in \Lambda \setminus \Lambda^+$. This is because for $\mu \in \Lambda \setminus \Lambda^+$ there exists a simple root $\mathbf{a}_j$ (2.2) for which $\langle \mathbf{a}_j, \mu \rangle$ is a negative integer and hence we pick up a zero from the factor

$$\frac{1}{(q^{1-g+\langle \mathbf{a}_j,\rho \rangle}; q)_{\langle \mathbf{a}_j,\mu \rangle}} = \frac{1}{(q;q)_{\langle \mathbf{a}_j,\mu \rangle}}$$

(which is zero for $\langle \mathbf{a}_j, \mu \rangle < 0$). The expressions for the r.h.s. are obtained from that of (A.1) (with $\mathbf{z} = \rho$) by canceling common factors in numerator and denominator. □

A further reduction arises when we specialize the parameters $q$ and $g$ in such a way that $q^{g(N+1)+M} = 1$ for some positive integer $M$. The sum of Proposition A.2 then terminates to a sum over the integral alcove $\Lambda_M^+$ (3.6).

**Proposition A.3** (A terminating Aomoto-Ito-Macdonald sum). *Let*

$$q = \exp\left(\frac{2\pi i}{g(N+1) + M}\right)$$

*with $g > 0$ and $M$ a positive integer. Then*

$$\sum_{\mu \in \Lambda_M^+} q^{-2\langle \rho,\mu \rangle} \prod_{\mathbf{a} \in A_N^+} \left(\frac{1 - q^{\langle \mathbf{a},\rho+\mu \rangle}}{1 - q^{\langle \mathbf{a},\rho \rangle}}\right) \frac{(q^{g+\langle \mathbf{a},\rho \rangle}; q)_{\langle \mathbf{a},\mu \rangle}}{(q^{1-g+\langle \mathbf{a},\rho \rangle}; q)_{\langle \mathbf{a},\mu \rangle}}$$

$$= (N+1) \prod_{1 \leq n \leq N} (q^{1+ng}; q)_{M-1}.$$

*Proof.* Let us first substitute

(A.3) $$g = -\frac{M}{N+1} + \frac{2\pi i}{(N+1)\log(q)}$$

with $0 < q < 1$ and $M$ a positive integer, in the summation formula of Proposition A.2. Notice that this value of $g$ satisfies both the convergence criterion $Re(g) < 0$ as well as the regularity condition that $g - \langle \mathbf{a}, \rho \rangle - 1 \notin \mathbb{N} + \frac{2\pi}{i \log(q)} \mathbb{Z}$ for all $\mathbf{a} \in A_N^+$. The sum over the dominant cone $\Lambda^+$ (2.7) then terminates to a sum over the integral alcove $\Lambda_M^+$ (3.6) because all terms become zero for $\mu \in \Lambda^+ \setminus \Lambda_M^+$. Indeed, for the above value of $g$ we have that $q^{(N+1)g+M} = 1$, so we pick up a zero from the factor

$$(q^{g+\langle \mathbf{a}_{max},\rho \rangle}; q)_{\langle \mathbf{a}_{max},\mu \rangle} = (q^{(N+1)g}; q)_{\mu_1 - \mu_{N+1}}$$

when $\langle \mathbf{a}_{max}, \mu \rangle = \mu_1 - \mu_{N+1} > M$. To arrive at the expression for the r.h.s. one uses that

$$\prod_{1 \leq n \leq N} \frac{(q^{1+ng}; q)_\infty}{(q^{-ng}; q)_\infty} = \prod_{1 \leq n \leq N} \frac{(q^{1+ng}; q)_\infty}{(q^{(N+1-n)g+M}; q)_\infty} = \prod_{1 \leq n \leq N} (q^{1+ng}; q)_{M-1}.$$

We thus see that the terminating sum of the proposition holds for $q$ given by $\exp(\frac{2\pi i}{(N+1)g+M})$ with $Re(g) = -M/(N+1)$ and $Im(g) < 0$ (solve (A.3) for $q$). By exploiting the analyticity in $g$ it is possible to extend the terminating sum to generic complex $g$. The restriction to positive real values of $g$ ensures that all numerators and denominators are nonzero. □

In trigonometric notation with $q = e^{i\alpha}$, the summation formula of Proposition A.3 becomes

$$
\text{(A.4)} \quad \sum_{\mu \in \Lambda_M^+} \prod_{\mathbf{a} \in A_N^+} \frac{\sin(\frac{\alpha}{2}\langle \mathbf{a}, \rho + \mu \rangle) \prod_{m=1}^{\langle \mathbf{a}, \mu \rangle} \sin \frac{\alpha}{2}(\langle \mathbf{a}, \rho \rangle + g + m - 1)}{\sin(\frac{\alpha}{2}\langle \mathbf{a}, \rho \rangle) \prod_{m=1}^{\langle \mathbf{a}, \mu \rangle} \sin \frac{\alpha}{2}(\langle \mathbf{a}, \rho \rangle - g + m)}
$$
$$
= 2^{N(M-1)}(N+1) \prod_{\substack{1 \leq m \leq M-1 \\ 1 \leq n \leq N}} \sin \frac{\alpha}{2}(m + ng)
$$

for $\alpha = 2\pi/(g(N+1) + M)$ with $M$ a positive integer and $g > 0$. (Here we have used the convention that empty products are equal to 1.) This is precisely the summation formula of Proposition 4.4.

*Remarks:* i. For $N = 1$ the Aomoto-Ito-Macdonald-type sums of Eq. (A.1), Proposition A.2 and Proposition A.3 reduce to

$$
\text{(A.5)} \quad \sum_{m=-\infty}^{\infty} q^{-gm} \Big( \frac{1 - q^{z+m}}{1 - q^z} \Big) \frac{(q^{g+z}; q)_m}{(q^{1-g+z}; q)_m} = 2 \frac{(q^{1+z}, q^{1-z}; q)_\infty}{(q^{1-g+z}, q^{1-g-z}; q)_\infty} \frac{(q^{1-2g}; q)_\infty}{(q^{1-g}; q)_\infty}
$$

(with $0 < q < 1$, $Re(g) < 0$ and $z \notin \frac{2\pi}{i \log(q)}\mathbb{Z}$, $g \pm z - 1 \notin \mathbb{N} + \frac{2\pi}{i \log(q)}\mathbb{Z}$),

$$
\text{(A.6)} \quad \sum_{m=0}^{\infty} q^{-gm} \Big( \frac{1 - q^{g+m}}{1 - q^g} \Big) \frac{(q^{2g}; q)_m}{(q; q)_m} = 2 \frac{(q^{1+g}; q)_\infty}{(q^{-g}; q)_\infty}
$$

(with $0 < q < 1$, $Re(g) < 0$) and

$$
\text{(A.7)} \quad \sum_{m=0}^{M} q^{-gm} \Big( \frac{1 - q^{g+m}}{1 - q^g} \Big) \frac{(q^{2g}; q)_m}{(q; q)_m} = 2(q^{1+g}; q)_{M-1}
$$

(with $q = \exp(\frac{\pi i}{g+M/2})$ and $g > 0$), respectively. The sums in (A.5), (A.6) and (A.7) are well-poised $_2\psi_2$, $_2\phi_1$ and terminating $_2\phi_1$ sums that arise as reductions of Bailey's $_6\psi_6$, Rogers' $_6\phi_5$ and Rogers' terminating $_6\phi_5$ very-well-poised sums, respectively [GR].

ii. The summation formula of Aomoto, Ito and Macdonald in [Ao, I, M5] is in fact more general than as stated in Proposition A.1. This is because they consider sums associated to an arbitrary reduced integral root system. The formulation of Proposition A.1 corresponds to the restriction of [Ao, I, M5] to case of the $A_N$ series. In [D3] a further extension of the Aomoto-Ito-Macdonald sum for the nonreduced $BC$-type root systems was studied together with corresponding truncated and terminating variants.

## Appendix B. Bilinear summation identities for Macdonald's symmetric functions

The purpose of this appendix is twofold. Firstly, it serves to collect some basic facts on the Macdonald symmetric functions that were needed in Sect 4. For a more complete treatment of this material and proofs the reader is referred to [M4, Ch. 6]. Secondly, the appendix allows us to reformulate some of our results from the perspective of algebraic combinatorics. This gives rise to a novel system of bilinear summation identities for the Macdonald symmetric functions (cf. Proposition B.2 below).

Let

$$\boldsymbol{D}_r = t^{r(r-1)/2} \sum_{\substack{J \subset \{1,\ldots,N+1\} \\ |J|=r}} \prod_{\substack{j \in J \\ k \notin J}} \left( \frac{tz_j - z_k}{z_j - z_k} \right) T_{J,q}, \qquad r = 1, \ldots, N+1, \tag{B.1}$$

where $|J|$ denotes the cardinality of the index set $J \subset \{1, \ldots, N+1\}$ and $T_{J,q} = \prod_{j \in J} T_{j,q}$ with

$$(T_{j,q} f)(z_1, \ldots, z_{N+1}) = f(z_1, \ldots, z_{j-1}, qz_j, z_{j+1}, \ldots, z_{N+1}).$$

Let $\mathbf{n} = (n_1, n_2, \ldots, n_{N+1}) \in \mathbb{N}^{N+1}$ be a partition, i.e., let the components (or parts) be ordered as $n_1 \geq n_2 \geq \cdots \geq n_{N+1} \geq 0$. The monomial symmetric function $\boldsymbol{m}_{\mathbf{n}}(\mathbf{z})$ associated to $\mathbf{n}$ is then defined as

$$\boldsymbol{m}_{\mathbf{n}}(\mathbf{z}) = \sum_{\mathbf{m} \in S_{N+1}(\mathbf{n})} z_1^{m_1} \cdots z_{N+1}^{m_{N+1}}, \tag{B.2}$$

where the sum is over the orbit of $\mathbf{n}$ under the action of the permutation group $S_{N+1}$ on the components. The basis of monomial symmetric functions $\{\boldsymbol{m}_{\mathbf{n}}\}$ inherits a partial order from the *dominance partial order* of the partitions defined by

$$\mathbf{m} \preceq \mathbf{n} \quad iff \quad |\mathbf{m}| = |\mathbf{n}| \quad and \quad m_1 + \cdots + m_k \leq n_1 + \cdots + n_k \tag{B.3}$$

for $k = 1, \ldots, N$, where $|\mathbf{n}| := n_1 + \cdots + n_{N+1}$ denotes the weight of the partition.

**Proposition B.1** (Triangularity [M4]). *The $q$-difference operators $D_1, \ldots, D_{N+1}$ commute and are triangular with respect to the basis of monomial symmetric functions*

$$\boldsymbol{D}_r \boldsymbol{m}_{\mathbf{n}} = \sum_{\mathbf{m} \preceq \mathbf{n}} [\boldsymbol{D}_r]_{\mathbf{n},\mathbf{m}} \boldsymbol{m}_{\mathbf{m}}, \qquad [\boldsymbol{D}_r]_{\mathbf{n},\mathbf{m}} \in \mathbb{Q}(q,t).$$

*Furthermore, the diagonal matrix elements (eigenvalues) $[\boldsymbol{D}_r]_{\mathbf{n},\mathbf{n}}$ are given by*

$$[\boldsymbol{D}_r]_{\mathbf{n},\mathbf{n}} = \boldsymbol{E}_{\mathbf{n},r}(q,t) = \sum_{J \subset \{1,\ldots,N+1\}} \prod_{j \in J} t^{N+1-j} q^{n_j}.$$

For a quick proof of the polynomiality of $(\boldsymbol{D}_r \boldsymbol{m}_{\mathbf{n}})(\mathbf{z})$ in $\mathbf{z}$ one may use that this rational expression is regular as a function of $\mathbf{z} \in \mathbb{C}^{N+1}$ due to the permutation symmetry. The triangular form of the monomial expansion for $(\boldsymbol{D}_r \boldsymbol{m}_{\mathbf{n}})(\mathbf{z})$ and the diagonal matrix elements $[\boldsymbol{D}_r]_{\mathbf{n},\mathbf{n}}$ then follow from the asymptotics for $\mathbf{z}$ to infinity.

In the simplest situation, i.e. for $\mathbf{n} = \mathbf{0}$ (so $\boldsymbol{m_n} = 1$), the monomial expansion of Proposition B.1 reduces to the ($A_N$-type) Macdonald identity (cf. [M1, Theorem (2.8)])

$$\text{(B.4)} \quad t^{r(r-1)/2} \sum_{\substack{J \subset \{1,\ldots,N+1\} \\ |J|=r}} \prod_{\substack{j \in J \\ k \notin J}} \left( \frac{tz_j - z_k}{z_j - z_k} \right) = \sum_{\substack{J \subset \{1,\ldots,N+1\} \\ |J|=r}} \prod_{j \in J} t^{N+1-j},$$

with $r = 1, \ldots, N+1$. Substitution of $t = e^{i\alpha g}$ and $z_j = e^{i\alpha x_j}$, $j = 1, \ldots, N+1$ produces upon division by $t^{\frac{rN}{2}}$ the Macdonald identity used in the proof of Proposition 4.3 for $r = 1, \ldots, N$ (the $(N+1)$-th identity in (B.4) is in fact trivial, as in that case the product on the l.h.s. becomes empty and one ends up with the equality $t^{N(N+1)/2} = t^N t^{N-1} \cdots t \cdot 1$).

The Macdonald symmetric functions are now defined as the (joint) eigenfunctions of the commuting operators $\boldsymbol{D}_1, \ldots, \boldsymbol{D}_{N+1}$. Such a definition makes sense because the eigenvalues $\boldsymbol{E}_{\mathbf{n},r}(q,t)$ are nondegenerate (and hence semisimple) over the field $\mathbb{Q}(q,t)$.

**Definition** (Macdonald symmetric functions). For a partition $\mathbf{n}$ in $\mathbb{N}^{N+1}$, the *Macdonald symmetric function* $\boldsymbol{p_n}(\mathbf{z})$ is the symmetric polynomial of the form

(i) $\boldsymbol{p_n}(\mathbf{z}) = \boldsymbol{m_n}(\mathbf{z}) + \sum_{\mathbf{m} \prec \mathbf{n}} c_{\mathbf{n},\mathbf{m}}(q,t) \, \boldsymbol{m_m}(\mathbf{z})$ with $c_{\mathbf{n},\mathbf{m}}(q,t) \in \mathbb{Q}(q,t)$

such that

(ii) $\boldsymbol{D}_r \boldsymbol{p_n} = \boldsymbol{E}_{\mathbf{n},r}(q,t) \, \boldsymbol{p_n}$, $r = 1, \ldots, N+1$.

Two important properties of the Macdonald symmetric functions are the *evaluation formula* (also referred to as the *specialization formula*) [M4, Ch. 6: Eqs. (6.11), (6.11')]

$$\text{(B.5a)} \quad \boldsymbol{p_n}(\tau) = t^{\sum_{j=1}^{N+1}(j-1)n_j} \prod_{1 \le j < k \le N+1} \frac{(t^{1+k-j}; q)_{n_j - n_k}}{(t^{k-j}; q)_{n_j - n_k}}$$

with $\tau = (t^N, t^{N-1}, \ldots, t, 1)$ and the *symmetry relation* [M4, Ch. 6: Eq. (6.6)]

$$\text{(B.5b)} \quad \boldsymbol{P_m}(\tau q^{\mathbf{n}}) = \boldsymbol{P_n}(\tau q^{\mathbf{m}})$$

with

$$\text{(B.6)} \quad \boldsymbol{P_n}(\mathbf{z}) = \boldsymbol{p_n}(\mathbf{z}) / \boldsymbol{p_n}(\tau)$$

and $\tau q^{\mathbf{n}} = (t^N q^{n_1}, t^{N-1} q^{n_2}, \ldots, t q^{n_N}, q^{n_{N+1}})$.

The Macdonald symmetric function $\boldsymbol{p_n}(\mathbf{z})$ is homogeneous of degree $|\mathbf{n}|$ in $\mathbf{z}$ and

$$\boldsymbol{p}_{\mathbf{n} + \mathbf{e}_1 + \cdots + \mathbf{e}_{N+1}}(\mathbf{z}) = (z_1 \cdots z_{N+1}) \boldsymbol{p_n}(\mathbf{z}).$$

Projection to a homogeneous function of degree zero in $\mathbf{z}$

$$\text{(B.7a)} \quad \mathbf{p}_\lambda(\mathbf{z}) = (z_1 \cdots z_{N+1})^{-|\mathbf{n}|/(N+1)} \, \boldsymbol{p_n}(\mathbf{z}),$$

$$\text{(B.7b)} \quad \lambda = \mathbf{n} - \frac{|\mathbf{n}|}{N+1}(\mathbf{e}_1 + \cdots + \mathbf{e}_{N+1})$$

gives rise to the *Macdonald polynomials* $\mathbf{p}_\lambda$, $\lambda \in \Lambda^+$ (2.7) associated to the root system $A_N$ [M2, M3]. The functions $\mathbf{p}_\lambda(\mathbf{z})$ (B.7a), (B.7b) are related to the trigonometric Macdonald polynomials $p_\lambda(\mathbf{x})$ of Sect. 4.2 via the trigonometric substitution

(B.8a) $$t = q^g, \qquad q = e^{i\alpha}$$

(B.8b) $$z_j = e^{i\alpha x_j}, \quad j = 1, \ldots, N+1.$$

More specifically, by substituting (B.8a), (B.8b) the functions $\mathbf{p}_\lambda(\mathbf{z})$ (B.7a), (B.7b) pass over to trigonometric polynomials $p_\lambda(\mathbf{x})$ of the form in (4.10a), (4.10b). (The difference equations in (4.10b) are equivalent to the $q$-difference equations for $\mathbf{p}_\lambda(\mathbf{z})$, originating from the $q$-difference equations $\boldsymbol{D}_r \boldsymbol{p}_\mathbf{n} = \boldsymbol{E}_{\mathbf{n},r}(q,t)\,\boldsymbol{p}_\mathbf{n}$ in the above definition of the Macdonald symmetric function $\boldsymbol{p}_\mathbf{n}(\mathbf{z})$, upon substitution of (B.8a), (B.8b)). The symmetry relations in (4.12) for the renormalized trigonometric Macdonald polynomials $P_\lambda(\mathbf{x})$ (4.11) are an immediate consequence of evaluation formula (B.5a) and the symmetry relation (B.5b). The real-valuedness of the expansion coefficients $c_{\lambda,\mu}$ in (4.10a) follows from the fact that the Macdonald symmetric functions $\boldsymbol{p}_\mathbf{n}(\mathbf{z})$ are invariant with respect to the parameter inversion $(q,t) \to (q^{-1}, t^{-1})$, see [M4, Ch 6: Eq. (4.14) (iv)].

Translating back the orthogonality relations from Sect. 4.2 leads us to the following system of bilinear summation identities for the Macdonald symmetric functions $\boldsymbol{p}_\mathbf{n}(\mathbf{z})$.

**Proposition B.2** (Bilinear summation identities). *Let $\mathbf{n}, \mathbf{k}$ be partitions in $\mathbb{N}^{N+1}$ with $n_1 - n_{N+1}, k_1 - k_{N+1} \leq M \in \mathbb{N} \setminus \{0\}$. Then we have that*

$$\sum_{\substack{M \geq m_1 \geq \cdots \geq m_{N+1}=0 \\ \mathbf{m} \in \mathbb{N}^{N+1}}} q^{-\frac{|\mathbf{m}|(|\mathbf{n}|-|\mathbf{k}|)}{N+1}} \boldsymbol{p}_\mathbf{n}(\tau q^\mathbf{m}) \boldsymbol{p}_\mathbf{k}(\tau^{-1} q^{-\mathbf{m}}) \boldsymbol{\Delta}(\mathbf{m})$$

$$= \begin{cases} 0 & \text{if } \mathbf{k} \neq \mathbf{n} \mod \mathbb{Z}(\mathbf{e}_1 + \cdots + \mathbf{e}_{N+1}) \\ t^{\frac{N(|\mathbf{n}|-|\mathbf{k}|)}{2}} \boldsymbol{\mathcal{N}_0} \boldsymbol{\mathcal{N}}(\mathbf{n}) & \text{if } \mathbf{k} = \mathbf{n} \mod \mathbb{Z}(\mathbf{e}_1 + \cdots + \mathbf{e}_{N+1}) \end{cases}$$

*as rational identity in $q^{\frac{1}{N+1}}$ and $t$ subject to relation $t^{N+1} q^M = 1$, where*

$$\tau^{\pm 1} q^{\pm \mathbf{m}} = (t^{\pm N} q^{\pm m_1}, t^{\pm(N-1)} q^{\pm m_2}, \ldots, t^{\pm 1} q^{\pm m_N}, q^{\pm m_{N+1}})$$

*and*

$$\boldsymbol{\Delta}(\mathbf{m}) = t^{-\sum_{j=1}^{N+1}(N+2-2j)m_j} \prod_{1 \leq j < k \leq N+1} \left(\frac{1-t^{k-j}q^{m_j-m_k}}{1-t^{k-j}}\right) \frac{(t^{1+k-j}; q)_{m_j-m_k}}{(qt^{k-j-1}; q)_{m_j-m_k}},$$

$$\boldsymbol{\mathcal{N}}(\mathbf{n}) = \prod_{1 \leq j < k \leq N+1} \frac{(t^{1+k-j}, qt^{k-j-1}; q)_{n_j-n_k}}{(t^{k-j}, qt^{k-j}; q)_{n_j-n_k}},$$

$$\boldsymbol{\mathcal{N}_0} = (N+1) \prod_{1 \leq j \leq N} (qt^j; q)_{M-1}.$$

*(For $\mathbf{k} = \mathbf{n}$ mod $\mathbb{Z}(\mathbf{e}_1 + \cdots + \mathbf{e}_{N+1})$ the relation is actually rational in $q$ itself.)*

*Proof.* First the projection formulas in (B.7a), (B.7b) are used to rewrite the stated bilinear summation identities in terms of the $A_N$-type Macdonald polynomials $\mathbf{p}_\lambda$, $\lambda \in \Lambda_M^+$. The trigonometric substitution in (B.8a), (B.8b) then leads us back to the discrete orthogonality relations (4.15a) for the trigonometric Macdonald polynomials $p_\lambda(\mathbf{x})$, $\lambda \in \Lambda_M^+$ of Sect. 4.2. This proves our bilinear summation formulas for the Macdonald symmetric functions $\boldsymbol{p}_\mathbf{n}(\mathbf{z})$ for $t = q^g$, $q = \exp(\frac{2\pi i}{(N+1)g+M})$ with $g > 0$. Analytic continuation in $g$ entails the formulation of the proposition. $\square$

If we specialize the formula of Proposition B.2 to the case that $\mathbf{n} = \mathbf{m} = \mathbf{0}$, then we arrive at the following rational identity in $q, t$ subject to the relation $t^{N+1}q^M = 1$

$$\text{(B.9)} \qquad \sum_{\substack{M \geq m_1 \geq \cdots \geq m_{N+1} = 0 \\ \mathbf{m} \in \mathbb{N}^{N+1}}} \boldsymbol{\Delta}(\mathbf{m}) = \boldsymbol{\mathcal{N}_0},$$

which amounts to the terminating Aomoto-Ito-Macdonald sum of Proposition A.3.

The complex conjugation relation in (4.17), for the trigonometric Macdonald polynomials $p_\lambda(\mathbf{x})$ of Sect. 4.2, translates to a corresponding relation for the Macdonald symmetric functions $\boldsymbol{p}_\mathbf{n}(\mathbf{z})$ defined above. Specifically, if we associate to a partition $\mathbf{n} \in \mathbb{N}^{N+1}$ a contragredient partition $\mathbf{n}^* \in \mathbb{N}^{N+1}$ with parts given by

$$\text{(B.10)} \qquad n_j^* = n_1 - n_{(N+2-j)}, \qquad j = 1, \ldots, N+1$$

(see Figure 4), then we have that

$$\text{(B.11)} \qquad \boldsymbol{p}_{\mathbf{n}^*}(\mathbf{z}) = (z_1 \cdots z_{N+1})^{\frac{|\mathbf{n}|+|\mathbf{n}^*|}{N+1}} \boldsymbol{p}_\mathbf{n}(\mathbf{z}^{-1}) \qquad \left(\frac{|\mathbf{n}|+|\mathbf{n}^*|}{N+1} = n_1\right),$$

where $\mathbf{z}^{-1} := (z_1^{-1}, \ldots, z_{N+1}^{-1})$. Observe that the mapping $\mathbf{n} \mapsto \mathbf{n}^*$ is involutive modulo $\mathbb{Z}(\mathbf{e}_1 + \cdots + \mathbf{e}_{N+1})$, i.e., the contragredient partition of $\mathbf{n}^*$ is equal to $\mathbf{n}$ up to a possible integer multiple of the vector $\mathbf{e}_1 + \cdots + \mathbf{e}_{N+1}$ (in case $n_{N+1} > 0$). The verification of (B.11) goes along lines very similar to the proof of Proposition B.2. First one uses the projective relation between the Macdonald symmetric functions $\boldsymbol{p}_\mathbf{n}$ and the $A_N$-type Macdonald polynomials $\mathbf{p}_\lambda$ in (B.7a), (B.7a), to conclude that after the trigonometric substitution (B.8a), (B.8b) the relation in (B.11) reduces to (4.17). Notice to this end that if $\lambda$ is the projection of a partition $\mathbf{n} \in \mathbb{N}^{N+1} \subset \mathbb{R}^{N+1}$ on the hyperplane $E$ (2.1) (cf. (B.7b)), then $\lambda^*$ (4.16b) amounts to the projection of the contragredient partition $\mathbf{n}^*$ (B.10) onto $E$. This proves Eq. (B.11) for $t = q^g$ with $g > 0$ and $q, z_j$ on the unit circle. Analytic continuation then entails that Eq. (B.11) holds identically as an equality that is polynomial in $\mathbf{z}$ and rational in $q, t$.

For Schur functions ($t = q$) the formula in (B.11) is well-known, see e.g. Stanley [St, Eq. (11)]; it expresses an equivalence between (the characters of) the irreducible representation of $SL(N+1, \mathbb{C})$ associated to the partition $\mathbf{n}^*$ and the representation contragredient to the one associated to $\mathbf{n}$. We have not been able to locate a reference for the property (B.11) applying to the general $(q, t)$-Macdonald symmetric functions, but most likely it was known already for this case too.

With the aid of (B.11), one rewrites the equality of Proposition B.2 in the form

$$\text{(B.12a)} \sum_{\substack{M \geq m_1 \geq \cdots \geq m_{N+1} = 0 \\ \mathbf{m} \in \mathbb{N}^{N+1}}} q^{-\frac{|\mathbf{m}|(|\mathbf{n}|+|\mathbf{k}|)}{N+1}} \boldsymbol{p}_\mathbf{n}(\tau q^\mathbf{m}) \boldsymbol{p}_\mathbf{k}(\tau q^\mathbf{m}) \boldsymbol{\Delta}(\mathbf{m})$$

$$= \begin{cases} 0 & \text{if } \mathbf{k} \neq \mathbf{n}^* \mod \mathbb{Z}(\mathbf{e}_1 + \cdots + \mathbf{e}_{N+1}) \\ t^{\frac{N(|\mathbf{n}|+|\mathbf{k}|)}{2}} \boldsymbol{\mathcal{N}_0} \boldsymbol{\mathcal{N}}(\mathbf{n}) & \text{if } \mathbf{k} = \mathbf{n}^* \mod \mathbb{Z}(\mathbf{e}_1 + \cdots + \mathbf{e}_{N+1}), \end{cases}$$

FIGURE 4. The contragredient partition $\mathbf{n}^*$ of a partition $\mathbf{n} \in \mathbb{N}^{N+1}$.

or equivalently in the normalization of (B.6) (cf. also (4.15b))

$$\text{(B.12b)} \quad \sum_{\substack{M \geq m_1 \geq \cdots \geq m_{N+1}=0 \\ \mathbf{m} \in \mathbb{N}^{N+1}}} q^{-\frac{|\mathbf{m}|(|\mathbf{n}|+|\mathbf{k}|)}{N+1}} \boldsymbol{P_n}(\tau q^{\mathbf{m}}) \boldsymbol{P_k}(\tau q^{\mathbf{m}}) \boldsymbol{\Delta}(\mathbf{m})$$

$$= \begin{cases} 0 & if \ \mathbf{k} \neq \mathbf{n}^* \mod \mathbb{Z}(\mathbf{e}_1 + \cdots + \mathbf{e}_{N+1}) \\ \frac{\boldsymbol{\mathcal{N}_0}}{\boldsymbol{\Delta}(\mathbf{n})} & if \ \mathbf{k} = \mathbf{n}^* \mod \mathbb{Z}(\mathbf{e}_1 + \cdots + \mathbf{e}_{N+1}), \end{cases}$$

as a rational identity in $t$ and $q^{\frac{1}{N+1}}$ (or even $q$ if $\mathbf{k}^* = \mathbf{n} \mod \mathbb{Z}(\mathbf{e}_1 + \cdots + \mathbf{e}_{N+1})$) subject to the relation $t^{N+1} q^M = 1$, where $\mathbf{n}$ and $\mathbf{k}$ are partitions in $\mathbb{N}^{N+1}$ with $n_1 - n_{N+1}, k_1 - k_{N+1} \leq M$. (To derive (B.12b) from (B.12a) one divides by $\boldsymbol{p_n}(\tau)\boldsymbol{p_k}(\tau)$ and uses that $t^{-N|\mathbf{n}|/2}\boldsymbol{p_n}(\tau) = t^{-N|\mathbf{n}^*|/2}\boldsymbol{p_{n^*}}(\tau)$ and that $t^{N|\mathbf{n}|}\boldsymbol{\mathcal{N}}(\mathbf{n})/(\boldsymbol{p_n}(\tau))^2 = 1/\boldsymbol{\Delta}(\mathbf{n})$.)

*Remarks:* i. In full generality, Macdonald defined his symmetric polynomials for an arbitrary integral root system [M3]. For the nonreduced $BC$ root systems, Koornwinder [Ko2] subsequently found a further generalization leading to a class of Askey-Wilson polynomials [GR, KS] in several variables that contains all Macdonald polynomials associated to the classical root systems as special cases (cf. [D1, Sect. 5]). In [DS] finite-dimensional discrete orthogonality properties of a type analogous to those described by Proposition B.2 were studied for Koornwinder's generalized $BC$ Askey-Wilson-Macdonald polynomials. In the case of the discrete orthogonality structure, the $BC$ polynomials in question may be viewed as a multivariate generalization of the well-known $q$-Racah polynomials introduced by Askey and Wilson [AS, GR, KS].

ii. For $t = q^g$ with $g$ a nonnegative integer, the condition $t^{N+1}q^M = 1$ implies that $q$ is a root of unity. In this special case the properties of the Macdonald polynomials were studied by Kirillov, Jr. and Cherednik [Ki, C2]. In particular, Kirillov, Jr. connects the Macdonald polynomials at issue with the representation theory of the quantum group (quantized enveloping algebra) $U_q(sl_{N+1})$ for $q$ a root of unity.

complex (Kähler) manifolds and to Igor Lutsenko for helping with the figures. Thanks are also due to François Ziegler and Anatol N. Kirillov for drawing our attention to Refs. [GS2] and [St], respectively.

Most of the results reported in this paper were obtained while the authors were visiting the Mathematical Sciences Research Institute (MSRI) in Berkeley, taking part in the Combinatorics Program on Representation Theory and Symmetric Functions (Spring, 1997). It is a pleasure to thank the organizers for their invitation and the institute for its hospitality. The research was supported in part by the Natural Sciences and Engineering Research Council (NSERC) of Canada, le Fonds pour la Formation de Chercheurs et l'Aide à la Recherche (FCAR) du Québec, and at the MSRI in part by NSF grant #DMS 9022140.

CENTRE DE RECHERCHES MATHÉMATIQUES, UNIVERSITÉ DE MONTRÉAL, C.P. 6128, SUCCURSALE CENTRE-VILLE, MONTRÉAL (QUÉBEC), H3C 3J7 CANADA
  *E-mail address*: `vandieje@CRM.UMontreal.CA`
  *E-mail address*: `vinet@CRM.UMontreal.CA`